\documentclass[aps,pra,twocolumn,showkeys,showpacs,amsmath,amssymb,amsfonts,floatfix,superscriptaddress,nofootinbib,10pt]{revtex4-1}

\usepackage{amsmath}
\usepackage{amsfonts}
\usepackage{bm}
\usepackage[colorlinks=true,citecolor=blue,linkcolor=blue,urlcolor=blue]{hyperref}
\usepackage{graphicx}

\usepackage{color}
\usepackage[dvipsnames]{xcolor} 
\usepackage[utf8]{inputenc}
\usepackage{bbold}
\usepackage[toc,titletoc]{appendix}

\pagecolor{white} 
\definecolor{sph}{rgb}{0.0588, 0.3216, 0.7294} 
\definecolor{ppk}{rgb}{1.0, 0.4549, 0.0902} 

\newcommand{\changes}[1]{{\color{black}{{#1}}}}

\global\long\def\p{\prime}
\global\long\def\ket#1{|#1\rangle}
\global\long\def\bra#1{\langle#1|}
\global\long\def\proj#1#2{|#1\rangle\langle#2|}

\global\long\def\tr{\mathrm{tr}}
\global\long\def\im{\imath}

\newcommand{\dg} {{\dagger}}
\newcommand{\pd} {{\phantom\dagger}}

\newcommand{\ci}[1] {{{c}_{#1}^{\pd}}}
\newcommand{\cid}[1] {{c}_{#1}^\dg}

\DeclareMathAlphabet\mathbfcal{OMS}{cmsy}{b}{n}

\newcommand{\bS} {\bm{\Sigma}}

\newcommand{\bH}{\bm{\bar{H}}}

\newcommand{\ql}{\mathcal{L}}
\newcommand{\qs}{\mathcal{S}}
\newcommand{\qr}{\mathcal{R}}

\renewcommand{\[}{\begin{equation}}
\renewcommand{\]}{\end{equation}}

\newcommand{\cm}{\mathbfcal{C}}

\newcommand{\CMinf}{\cm^\infty}

\newcommand{\qd}{\mathcal{D}}

\newcommand{\NS}{N_\qs}

\newcommand{\Rel}{\qd_{\rm R}}
\newcommand{\Dep}{\qd_{\rm D}}
\newcommand{\RelCM}{\bar{\qd}_{\rm R}}
\newcommand{\DepCM}{\bar{\qd}_{\rm D}}
\newcommand{\gam}{\boldsymbol\gamma}
\newcommand{\sig}{\boldsymbol\sigma}
\newcommand{\Diagsig}{{\rm diag}(\sig)}
\newcommand{\gi}{m}                     
\newcommand{\gip}{m^\p}                 
\newcommand{\nsig}{N_\sigma}
\newcommand{\subc}{{\bm c}^\infty} 
\newcommand{\subd}{{\bm d}} 
\newcommand{\subcgam}{{\bm c}^\infty_\gamma} 
\newcommand{\A}{\mathbfcal{H}}
\newcommand{\Q}{\mathbfcal{Z}}
\newcommand{\UL}{\mathbfcal{W}}
\newcommand{\UR}{\mathbfcal{V}}
\newcommand{\D}{\boldsymbol\lambda}
\newcommand{\dDi}{\boldsymbol\Delta\boldsymbol\lambda^{-1}}	
\newcommand{\SO}{\mathbf{D}} 
\newcommand{\Ss}{\check{\qs}}
\newcommand{\Ls}{\hat{\qs}}

\newcommand{\RL}[2]{%
  \mathrel{\mathop{%
    \vcenter{\offinterlineskip
      \ialign{\hfil##\hfil\cr
        \hphantom{$\scriptstyle\mspace{8mu}{#1}\mspace{8mu}$}\cr
        \rightarrowfill\cr
        \vrule height0pt width 2em\cr
        \leftarrowfill\cr
        \hphantom{$\scriptstyle\mspace{8mu}{#2}\mspace{8mu}$}\cr
        \noalign{\kern-0.3ex}
      }%
    }%
  }\limits^{#1}_{#2}}%
}

\usepackage{tikz}
\usepackage{pgfplots}
\pgfplotsset{compat=newest}
\usetikzlibrary{plotmarks}
\usetikzlibrary{arrows.meta}
\usepgfplotslibrary{patchplots}
\usepackage{grffile}

\definecolor{mycolor1}{rgb}{0,0,1}%
\definecolor{mycolor2}{rgb}{0.1,0.2,0.7}%
\definecolor{mycolor3}{rgb}{0.2,0.4,0.4}%
\definecolor{mycolor4}{rgb}{0.33,0.33,0.33}%
\definecolor{mycolor5}{rgb}{0.4,0.4,0.2}%
\definecolor{mycolor6}{rgb}{0.7,0.2,0.1}%
\definecolor{mycolor7}{rgb}{1,0,0}%

\begin{document}
\title{Approaching the scaling limit of transport through lattices with dephasing}

\author{Subhajit Sarkar}
\affiliation{Institute of Theoretical Physics, Jagiellonian University, Łojasiewicza 11, 30-348 Kraków, Poland}
\affiliation{Biophysical and Biomedical Measurement Group, Microsystems and Nanotechnology Division, Physical Measurement Laboratory, National Institute of Standards and Technology, Gaithersburg, Maryland 20899, USA}
\affiliation{Department of Physics, School of Natural Sciences, Shiv Nadar Institution of Eminence Deemed to be University, Delhi-NCR, NH91, Tehsil Dadri, Greater Noida, Uttar Pradesh 201314, India}
\affiliation{Department of Physics and Nanotechnology, SRM Institute of Science and Technology, Kattankulathur-603 203, India.}

\author{Gabriela W\'ojtowicz}
\affiliation{Institute of Theoretical Physics, Jagiellonian University, Łojasiewicza 11, 30-348 Kraków, Poland}
\affiliation{Biophysical and Biomedical Measurement Group, Microsystems and Nanotechnology Division, Physical Measurement Laboratory, National Institute of Standards and Technology, Gaithersburg, Maryland 20899, USA}
\affiliation{Doctoral School of Exact and Natural Sciences, Jagiellonian University, Łojasiewicza 11, 30-348 Kraków, Poland}
\affiliation{Institute of Theoretical Physics, Albert-Einstein Allee 11, University of Ulm, D-89081 Ulm, Germany }
\author{Bart\l{}omiej Gardas}
\affiliation{Institute of Theoretical and Applied Informatics, Polish Academy of Sciences, Ba{\l}tycka 5, 44-100 Gliwice, Poland}
\author{Marek M. Rams}
\email{marek.rams@uj.edu.pl}
\affiliation{Institute of Theoretical Physics, Jagiellonian University, Łojasiewicza 11, 30-348 Kraków, Poland}
\affiliation{Mark Kac Complex Systems Research Center, Jagiellonian University, Łojasiewicza 11, 30-348 Kraków, Poland}
\author{Michael Zwolak}
\email{mpzwolak@gmail.com}
\affiliation{Biophysical and Biomedical Measurement Group, Microsystems and Nanotechnology Division, Physical Measurement Laboratory, National Institute of Standards and Technology, Gaithersburg, Maryland 20899, USA}

\begin{abstract}
We examine the stationary--state equations for lattices with generalized Markovian dephasing and relaxation. 
When the Hamiltonian is quadratic, the single--particle correlation matrix has a closed system of equations even in the presence of these two processes. 
The resulting equations have a vectorized form related to, but distinct from, Lyapunov's equation. 
We present an efficient solution that helps to achieve the scaling limit, e.g., of the current decay with lattice length. 
As an example, we study the super--diffusive--to--diffusive transition in a lattice with long--range hopping and dephasing. 
The approach enables calculations with up to $10^4$ sites, representing an increase of $10$ to $40$ times over prior studies. 
This enables a more precise extraction of the diffusion exponent, enhances agreement with theoretical results, and supports the presence of a phase transition. 
There is a wide range of problems that have Markovian relaxation, noise, and driving. 
They include quantum networks for machine--learning--based classification and extended reservoir approaches (ERAs) for transport. 
The results here will be useful for these classes of problems. 
\end{abstract}

\maketitle

\section{Introduction}
\label{sec:Intro}

Computational techniques for quantum transport are currently experiencing a renaissance, thanks to advances in many-body methods, such as tensor networks, and fundamental developments that numerically capture out--of--equilibrium systems. 
In the case of transport through junctions, the works of Subotnik, Hansen, Ratner, and Nitzan~\cite{subotnik_nonequilibrium_2009} and, independently, S\'anchez and collaborators~\cite{sanchez_molecular_2006,mceniry_dynamical_2007} led the development of relaxation--based methods. 
The incorporation of relaxation in ``extended,'' ``mesoscopic,'' or ``auxiliary'' reservoirs maintains a particle or temperature imbalance. 
This persistent imbalance leads to the formation of a genuine stationary state, in contrast to microcanonical approaches~\cite{cazalilla_time-dependent_2002,zwolak_mixed-state_2004,ventra_transport_2004,bohr_dmrg_2005,bushong_approach_2005,gobert_real-time_2005,al-hassanieh_adaptive_2006,cheng_simulating_2006,schneider_conductance_2006,schmitteckert_signal_2006,sai_microscopic_2007,bohr_strong_2007,zwolak_finite_2008,dias_da_silva_transport_2008,heidrich-meisner_real-time_2009,branschadel_conductance_2010,chien_bosonic_2012,chien_interaction-induced_2013,chien_landauer_2014,zwolak_communication_2018,gruss_energy-resolved_2018}.

\begin{figure}[t!]
    \centering
    \includegraphics[width=\columnwidth]{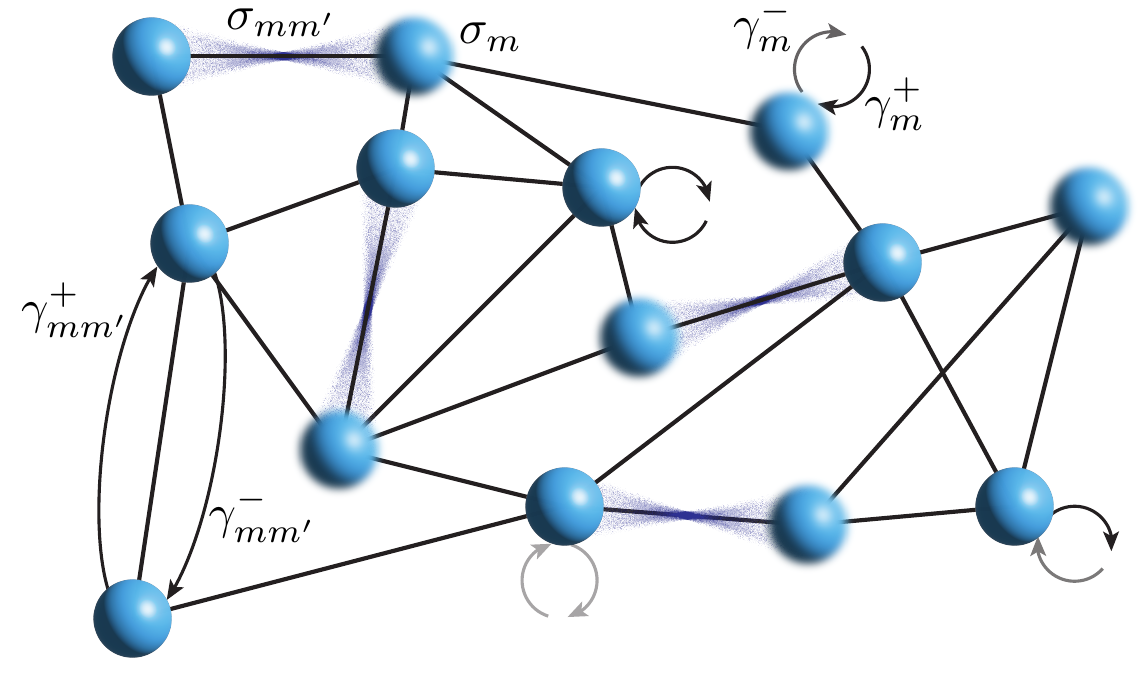}
    \caption{{\bf Markovian network dynamics.} 
    A network of modes undergo Markovian relaxation, represented with arrows and weights $\gamma^{+(-)}_{\gi\gip}$ ($\gamma^{+(-)}_{\gi}$ for onsite), and dephasing, represented by blurred bonds with strength $\sigma_{\gi\gip}$ or blurred sites with strength $\sigma_\gi$. 
    These processes drive dynamics and transport. 
    Extended reservoirs are an example of such a network where a subset of modes forms the left reservoir ($\ql$) and another the right ($\qr$), each with relaxation that maintains a potential or temperature drop. 
    Another example is boundary transport with injection and depletion only at terminal sites.}
    \label{fig:schematic}
\end{figure}

The pioneering works of Subotnik et al.~\cite{subotnik_nonequilibrium_2009} and S\'anchez et al.~\cite{sanchez_molecular_2006,mceniry_dynamical_2007} sparked numerous research threads, including the driven Liouville--von Neumann approach for single--particle dynamics~\cite{zelovich_state_2014,zelovich_moleculelead_2015, zelovich_driven_2016, hod_driven_2016, zelovich_parameter-free_2017,hod_driven_2023} and a Meir--Wingreen approach with Markovian relaxation for many--body systems that limits to the exact, continuum result~\cite{gruss_landauers_2016, gruss_communication_2017, elenewski_communication_2017, gruss_graphene_2018, zwolak_analytic_2020,zwolak_comment_2020, elenewski_performance_2021, wojtowicz_dual_2021, wojtowicz_accumulative_2023}.
This is among other formulations for electron transport~\cite{dzhioev_super-fermion_2011,ajisaka_nonequilibrium_2012, ajisaka_nonequilibrium_2013,chen_simple_2014,schwarz_lindblad-driven_2016,chiang_quantum_2020,chen_auxiliary_2019,chen_markovian_2019}, thermal transport~\cite{velizhanin_driving_2011,chien_tunable_2013,chien_thermal_2017,chien_topological_2018}, quantum thermodynamics~\cite{lacerda_entropy_2024}, reaction coordinates~\cite{binder_reaction_2018}, and relaxing~\cite{dast_quantum_2014,ajisaka_molecular_2015,mahajan_entanglement_2016,zanoci_entanglement_2020},  bosonic~\cite{tamascelli_nonperturbative_2018} systems, and pseudo--modes~\cite{imamoglu_stochastic_1994,garraway_nonperturbative_1997,garraway_decay_1997,zwolak_dynamics_2008,pleasance_generalized_2020}. 
Markovian processes are numerically efficient and compatible with various computational techniques, leading to several integration efforts. 
This includes density functional theory~\cite{morzan_electron_2017,hod_driven_2023,oz_electron_2023} and many--body methods. 
The latter encompass non--equilibrium impurity solvers~\cite{arrigoni_nonequilibrium_2013,dorda_auxiliary_2014,dorda_optimized_2017}, including with tensor networks~\cite{dorda_auxiliary_2015,fugger_nonequilibrium_2018,fugger_nonequilibrium_2020}. 
There are other tensor network implementations~\cite{lotem_renormalized_2020, brenes_tensor-network_2020}, including scalable mixed--basis tensor networks~\cite{rams_breaking_2020, wojtowicz_open-system_2020}. 
These leverage matrix product operators to describe open quantum systems~\cite{zwolak_mixed-state_2004,verstraete_matrix_2004}. 
In addition to stationary states, Floquet states are also being examined~\cite{de_transport_2023}, including with many--body interactions~\cite{de_confluence_2024} and for time crystals~\cite{sarkar_emergence_2022,sarkar_signatures_2022}. 

The challenges in this field are myriad. 
First and foremost, even for non--interacting systems, Markovian relaxation leads to unphysical behavior---most significantly, a violation of the fluctuation--dissipation theorem (FDT)~\cite{gruss_landauers_2016,elenewski_communication_2017,zwolak_analytic_2020}---and other anomalous behavior that hinders convergence to the true continuum reservoir~\cite{gruss_landauers_2016, gruss_communication_2017, elenewski_communication_2017, chiang_quantum_2020, zwolak_analytic_2020, elenewski_performance_2021, wojtowicz_dual_2021, wojtowicz_accumulative_2023, de_transport_2023}. 
Methods that employ Markovian relaxation to mimic a continuum should employ a limiting process: As the number of reservoir modes increases and the relaxation strength decreases, the reservoir description converges to the continuum. 
The exact choice of relaxation strength should ensure that the junction state, including observables such as the current, are stable~\cite{de_transport_2023} and away from anomalous regimes~\cite{wojtowicz_dual_2021,wojtowicz_accumulative_2023}. 
Due to this limit, the challenge will be to demonstrate the utility of Markovian relaxation in capturing continuum reservoirs (or thermal baths) in many--body scenarios, such as the Kondo problem, systems with phase transitions, or otherwise ones with a small energy scale. 
Moreover, the various threads should be woven together to identify the most efficient and stable approaches, especially when integrated with other techniques. 

The resulting methods, though, are already helpful for studying dynamics and transport on networks, see Fig.~\ref{fig:schematic}. 
Non--equilibrium stationary states under boundary--driven transport---a specific case of Markovian relaxation---and dephasing display a host of interesting phenomena, from rectification to phase transitions~\cite{landi_nonequilibrium_2022}. 
In that vein, we will provide a formulation of generalized Markovian dephasing and relaxation for systems with quadratic fermionic Hamiltonians. 
A cost analysis shows how to efficiently solve these equations. 
We demonstrate that this approach helps achieve the scaling limit, more accurately determine decay exponents, and locate the critical point in a one--dimensional lattice with onsite dephasing and long--range hopping~\cite{dhawan_anomalous_2024,sarkar_impact_2024,tater_bipartite_2025}.

\section{Equation of motion}
\label{sec:EOM}

We start from the equation of motion for the many--body density matrix, $\rho$, in the {\em Gorini--Kossakowski--Sudarshan--Lindblad} form~\cite{lindblad_generators_1976, gorini_completely_1976, breuer_theory_2007} 
\[ \label{eq:MarkovianERA}
\dot{\rho} = - \im [H, \rho] + \Rel [ \rho ]{+\Dep [ \rho ] } .
\]
Here, $H$ is the Hamiltonian, and $\Rel$ and $\Dep$ are relaxation and dephasing superoperators, respectively. 
We consider quadratic---non--interacting---Hamiltonians,
\[ \label{eq:HNonInt}
H = \sum_{\gi,\gip} \bH_{\gi\gip} \cid{\gi} \ci{\gip} ,
\]
that are particle--conserving, where $\bH_{\gi\gip}$ are the hopping coefficients and $\cid{\gi}$ ($\ci{\gip}$) are fermionic creation (annihilation) operators for mode $m$. 
This class of Hamiltonians does not have electron--electron or electron--phonon interactions. 
The Markovian relaxation is also quadratic, 
\begin{align} \label{eq:Rel}
\Rel[ \rho ] =
    & \sum_{\gi\gip}  \gamma^+_{\gi\gip} \left( \cid{\gi} \rho \ci{\gip} - \frac{1}{2} \left \{ \ci{\gip} \cid{\gi}, \rho\right \}\right) \notag \\
    + & \sum_{\gi\gip} \gamma^-_{\gi\gip} \left( \ci{\gi} \rho \cid{\gip} - \frac{1}{2} \left \{ \cid{\gip} \ci{\gi}, \rho \right \} \right) ,
\end{align}
where $\gamma^{+(-)}_{\gi\gip}$ are generalized injection (depletion) rates and $\{ \cdot,\cdot \}$ gives the anticommutator. 
This form appears in the auxiliary master equation approach~\cite{arrigoni_nonequilibrium_2013,dorda_auxiliary_2014}. 
For onsite relaxation of others~\cite{gruss_landauers_2016,brenes_tensor-network_2020}, we denote $\gamma^{+(-)}_{\gi\gi} \equiv \gamma^{+(-)}_{\gi}$. 
The Hermitian, positive semidefinite matrices, 
\[ \label{eq:gamma}
\gam^{+(-)} = \sum_{\gi\gip} \gamma^{+(-)}_{\gi\gip} \ket{\gi}\bra{\gip},
\]
completely describe the injection and depletion of fermions, and $\gam = \gam^+ + \gam^-$ is a matrix that gives the total dissipation strength. 
This form of relaxation, Eq.~\eqref{eq:Rel}, has a simple interpretation as the Markovian limit of linear contact with non--interacting fermionic reservoirs. 
This limit, though, requires multiple completely full and empty infinite--bandwidth reservoirs, which is the source of breaking the FDT~\cite{elenewski_communication_2017}.

Unlike relaxation, the dephasing is quartic, 
\[ \label{eq:Dep}
\Dep[\rho] = \sum_{\gi\gip} \sigma_{\gi\gip} \left( n_\gi \rho n_{\gip} - \frac{1}{2} \left \{ n_{\gip} n_\gi , \rho\right \} \right) ,
\]
where $n_\gi = \cid{\gi} \ci{\gi}$ is particle--number operator for mode $\gi$. 
The superoperator in Eq.~\eqref{eq:Dep} is a generalized form of dephasing that contains cross--terms between modes $\gi$ and $\gip$. 
The coefficient matrix,
\[ \label{eq:sig}
\sig = \sum_{\gi\gip} \sigma_{\gi\gip} \ket{\gi}\bra{\gip},
\]
is a real, symmetric, positive semidefinite matrix. 
The interpretation of this is straightforward: When an external environment interacts with a total, or subtotal, of particles---for instance, a capacitive coupling to the electron number on a quantum dot---one obtains such cross terms.
For local dephasing, i.e.,  $\sigma_{\gi\gip}=\sigma_{\gi}\delta_{\gi\gip}$ where $\delta_{\gi\gip}$ is the Kronecker delta function, the operator decoheres each mode $\gi$ independently. 

We focus on observables quadratic in fermionic operators, such as populations and particle current. 
To do so, we examine the correlation matrix, 
\[ \label{eq:CM}
\cm_{\gi\gip} = \tr [ \cid{\gip} \ci{\gi} \rho ] .
\]
Dephasing, though, is not a Gaussian process, i.e., Wick's theorem is not obeyed. 
Yet, Eq.~\eqref{eq:MarkovianERA} gives a closed system of equations for $\cm$ under local dephasing~\cite{landi_nonequilibrium_2022}. 
This holds also for generalized dephasing, Eq.~\eqref{eq:Dep}, yielding
\[ \label{eq:EOMcm}
\dot{\cm} = -\im \left[\bH, \cm \right] + \RelCM [ \cm ]+ \DepCM [ \cm ] ,
\]
with the superoperator
\[ \label{eq:RelCM}
\RelCM [ \cm ] = \gam^+ - \left\{ \gam , \cm \right\}/ 2 
\]
for relaxation~\cite{elenewski_communication_2017} and
\[ \label{eq:DepCM}
    \DepCM [\cm] = \sig \odot \cm - \left \{ \Diagsig, \cm \right \}/2 
\]
for dephasing. 
The $\odot$ signifies the Hadamard product. 

In the absence of dephasing, Eq.~\eqref{eq:EOMcm} is a Lyapunov equation, which gives a numerically efficient approach---scaling as $N^3$ with $N$ total modes in the network---to find the dynamics and stationary states, as well as to examine stability. 
As noted in Ref.~\cite{landi_nonequilibrium_2022}, common numerical routines~\cite{bartels_algorithm_1972,golub_hessenberg-schur_1979} for the Lyapunov equation can reach order $10^4$ sites at $N^3$ cost scaling. 
When dephasing is present, the resulting equation of motion is no longer in Lyapunov form.
This comes with a penalty. 
A general solution of Eq.~\eqref{eq:EOMcm} requires vectorization and diagonalization of the associated Lindblad superoperator, at a cost of $N^6$. 
For special forms of dephasing, one expects better scaling. 
We analyze the cost for generalized dephasing, as well as both local dephasing and dephasing only in a junction region. 
Local dephasing comes with a cost of $N^4$. 
Despite the additional power of $N$ compared to the Lyapunov equation---$N^4$ compared to $N^3$---we show that one, surprisingly, can still reach $N=10^4$ sites in reasonable compute time (several hours) on a single machine.

\section{Steady--state solution}
\label{sec:SS}

Defining the non--Hermitian Hamiltonian,
\[ \label{eq:A}
\A= - \im \bH - \gam/2 - \Diagsig/2 ,
\]
and the particle source, 
\[ \label{eq:Q}
\Q = \gam^+ ,
\]
the stationary state has the solution~\cite{turkeshi_diffusion_2021,sarkar_impact_2024}
\[ \label{eq:StationaryStateIntegral}
\CMinf = \int_0^\infty d\tau e^{-\im \A \tau} \bS^< e^{\im \A \tau} ,
\] 
where $\bS^< = \Q + \sig \odot \CMinf \equiv \bS^<_\gamma +  \bS^<_\sigma$. 
The appearance of $\CMinf$ on the left-- and right--hand sides makes the solution self--consistent and is a consequence of many--body correlations. 
While $\bS^<$ is just playing the role of a functional of $\CMinf$, it is, as the symbol implies, the lesser self--energy from non--equilibrium Green's functions. 

For the non--Hermitian Hamiltonian, Eq.~\eqref{eq:A}, we denote $\UL^\dg$ as the matrix with left eigenvectors as rows, $\UR$ with right eigenvectors as columns, and normalization $\UL^\dg \UR = I$ and eigenvalues $\D$. 
Equation~\eqref{eq:StationaryStateIntegral} becomes
\[ 
\CMinf = -\im \int \frac{d\omega}{2 \pi}  \UR  \frac{1}{\omega - \D} \tilde{\bS}^<  \frac{1}{\omega - \D^\star} \UR^\dg ,
\]
where $\tilde{\bS}^< = \UL^\dg \bS^< \UL$. 
This integration yields 
\[ \label{eq:ClosedSolution}
\CMinf = \UR \left[ \dDi \odot \tilde{\bS}^< \right] \UR^\dg ,
\]
with $\dDi$ having elements
\[
\left[ \dDi \right]_{pq} = 1/\left( \lambda_q^\star - \lambda_p \right) .
\] 
Equation~\eqref{eq:ClosedSolution} is linear in the two terms, so we can write
\[\label{eq:solution}
\begin{split}
\CMinf &= \UR \left[ \dDi \odot \UL^\dg \left( \bS^<_\gamma +  \bS^<_\sigma \right) \UL \right] \UR^\dg \\
&\equiv \CMinf_\gamma + \CMinf_\sigma ,
\end{split}
\]where the first term is independent of $\CMinf$ and the second, 
\[
\CMinf_\sigma = \im \UR \left[ \dDi \odot \left( \UL^\dg \left[ \sig \odot \CMinf \right] \UL \right) \right] \UR^\dg ,
\]
depends on $\CMinf$. 
The total solution thus requires solving, 
\[ \label{eq:CMinf}
\CMinf = \CMinf_\gamma + \SO \left[ \CMinf \right] ,
\]
with the elements of the superoperator $\SO$, 
\[ \label{eq:Delements}
\SO_{o o^\p, \gi \gip} = \im \sigma_{\gi\gip} \left[\UR \left[ \dDi \odot \UL^\dg \proj{\gi}{\gip} \UL \right] \UR^\dg\right]_{o o^\p} .
\]
Both Eq.~\eqref{eq:CMinf} and Eq.~\eqref{eq:Delements} can be costly operations.

\section{Computational Cost}
\label{sec:Cost}

The general solution of Eq.~\eqref{eq:CMinf} has a computational cost of $N^6$ to solve the vectorized matrix equation, 
\[ \label{eq:FullSolution}
\CMinf = ({\bm 1} - \SO)^{-1} \left[ \CMinf_\gamma \right],
\]
where ${\bm 1}$ is the identity matrix. 
There is a cost scaling of $N^5$ to find all the elements of $\SO$.
The lower power to compute all the elements exploits that for each of the $N^2$ pairs $\gi \gip$, the inner bracket in Eq.~\eqref{eq:Delements} requires $N^2$ operations. 
Sequentially, the outer bracket can be computed for all elements $o o^\p$ at once via matrix multiplication, for a cost of $N^3$. 
The total cost is thus $N^2 \left( N^2 + N^3 \right)$. 
If one were to compute the elements for each quadruple index $o o^\p \gi \gip$ one by one, then the cost would be $N^6$. 

Turkeshi and Schir\'o have a similar solution for a nearest--neighbor hopping model with onsite dephasing and boundary injection and depletion~\cite{turkeshi_diffusion_2021}, and analogous results appear in Refs.~\cite{varma_fractality_2017,taylor_subdiffusion_2021}. 
The model they study exhibits diffusive transport with resistivity that scales with the length of the lattice. 
Although they note that not all terms in the linear equations are necessary to find the density and current, their example problem does not require one to reach very large sizes in order to identify criticality or a crossover, as no such transition occurs in their model.
Moreover, as we will see via a full cost analysis, computing all elements of $\CMinf$ comes at no additional cost in terms of scaling with network size.  

Equations~\eqref{eq:MarkovianERA}--\eqref{eq:Dep} provide a more general form of relaxation and dephasing, including cross terms for both, as well as injection and depletion on an arbitrary number of sites (e.g., encompassing extended reservoir approaches). 
The solution, Eq.~\eqref{eq:CMinf} and Eq.~\eqref{eq:Delements}, should be solved sequentially: 
The terms of $\CMinf$ that correspond to non--zero elements of $\sig$ should be solved first, and then the remaining elements of $\CMinf$ can be found from the Lyapunov solution, Eq.~\eqref{eq:ClosedSolution}. 
Below, we will provide a complete cost analysis of this approach.
This solution enables simulations for system sizes that are 10 to 40 times larger than prior works~\cite{turkeshi_diffusion_2021,sarkar_impact_2024,taylor_subdiffusion_2021}, and on par with sizes achievable when Lyapunov solvers are applicable, despite the increase to $N^4$ scaling.
It allows us to extract the critical point more precisely for a model exhibiting a super--diffusive--to--diffusive transition~\cite{sarkar_impact_2024}. 

Considering the dephasing matrix $\sig$, let $\nsig$ be the number of non--zero elements, $0 \le \nsig \le N^2$.
For each non--zero element, one has to perform the matrix multiplications in Eq.~\eqref{eq:Delements} for the $N$ modes of the network. 
The computational cost to compute every element of $\SO$ is $\nsig N^3$.
Yet, solving Eq.~\eqref{eq:CMinf} sequentially, first for the elements of $\CMinf$ for which $\sig$ is non--zero, then for the rest of $\CMinf$, requires only the $\nsig^2$ elements of $\SO$ that act on the $\nsig$ subspace. 
Denoting quantities restricted to this subspace as lower case---$\subc$, $\subd$, $\subcgam$---the solution is
\[ \label{eq:SubSolution}
\subc = ({\bm 1} - \subd)^{-1} \left[ \subcgam \right] .
\]
To obtain the whole $\CMinf$, one forms $\sig \odot \CMinf$, which only requires $\subc$ since $\sig$ is zero on other elements.  
With $\subc$ in hand, Eq.~\eqref{eq:ClosedSolution} has only a cost of $N^3$ to obtain $\CMinf$.

The overall cost thus depends on the exact value of $\nsig$ relative to the number of total modes $N$.  
The cost is 
\[ \label{eq:Cost}
\mathrm{Cost} = \max \left[ \min \left( \nsig^2 N^2, N^5 \right), \nsig^3, N^3  \right] .
\]
From left to right, these are the cost of computing the $\nsig^2$ elements (cost, $\nsig^2 N^2$) or all elements (cost, $N^5$) of $\SO$, the cost to invert $({\bm 1}-\subd)$ in Eq.~\eqref{eq:SubSolution}, and the cost of all standard matrix operations. 
The latter includes applying Eq.~\eqref{eq:ClosedSolution} when $\subc$ is already known, and also the eigenvalue problem of the non--Hermitian Hamiltonian, Eq.~\eqref{eq:A}, itself. 
The $\nsig^2 N^2$ cost for the $\nsig^2$ elements of $\SO$ computes them term by term.
This limits to $N^6$ for a dense $\sig$. 
The appearance of $\min \left( \nsig^2 N^2, N^5 \right)$ in the cost indicates that when $\nsig$ is sufficiently large (i.e., when $\nsig^2 \ge N^3$), it is more efficient to compute all elements of $\SO$. 
The formal scaling still indicates that employing Eq.~\eqref{eq:SubSolution} is more efficient than Eq.~\eqref{eq:FullSolution} for any allowed $\nsig$, even though it is more efficient to compute all elements of $\SO$ in the formation step.

One of the most common cases is the presence of local dephasing at every site---i.e., $\sigma_{\gi\gip} \propto \delta_{\gi\gip}$. 
This requires first solving for only the stationary state occupations, ${\bm n}_m = \CMinf_{mm}$ with dephasing and ${\bm n}_{\gamma m} = \CMinf_{\gamma \, mm}$ with only relaxation.
The resulting equation is
\[
{\bm n} = ({\bm 1} - \subd)^{-1} \left[ {\bm n}_\gamma \right] .
\]
The cost, Eq.~\eqref{eq:Cost}, is thus
\[ \label{eq:CostOnsiteAll}
\mathrm{Cost} \to \max \left[ \min \left( N^4, N^5 \right), N^3, N^3  \right] = N^4 .
\]
Thus, it is the formation of $\subd$ that sets the cost. 
We note that this task is trivially parallelizable. 
The matrices $\UR$, $\UL$, $\dDi$, and $\sig$ can be distributed across many compute cores or nodes, where each calculates some subset of elements. 
When formation of the superoperator elements is the dominant cost, therefore, further gains in efficiency can be obtained, which we do numerically exploit. 
Importantly, there is no increase in memory requirements in forming $\subd$ for local dephasing, it remains at $N^2$. 

There are many other physically interesting cases, each potentially having a different dominant cost.  
For instance, transport across a junction region, $\qs$, driven by bias or temperature drop across non--Markovian reservoirs $\ql$ and $\qr$ is one of the most important classes. 
This setup requires taking the number of extended reservoir modes $N_r$ to infinity while taking the relaxation to zero. 
The junction size, though, remains either fixed or much smaller than $N_r$, i.e., $N_\qs \ll N_r$ with $N_\qs$ the number of sites in $\qs$. 
When there is local dephasing on just $\qs$, e.g., due to local inelastic scattering or electron--electron interactions, then nearly all of the elements of Eq.~\eqref{eq:Delements} are zero. 
In this case, Eq.~\eqref{eq:Cost} is 
\[ \label{eq:CostDepOnS}
\mathrm{Cost} \to \max \left[ \min \left( N_\qs^2 N^2, N^5 \right), N_\qs^3, N^3  \right] = N^3 .
\]
Here, $N=2 N_r + N_\qs$ and, since $N_r \to N_\qs$ at fixed $N_\qs$, we assume $N_\qs^2 \le N$. 
That is, the cost scaling for this setup is the same as without dephasing.

\section{Example}
\label{sec:Example}

\begin{figure}[t!]
    \centering
    \includegraphics[width=\columnwidth]{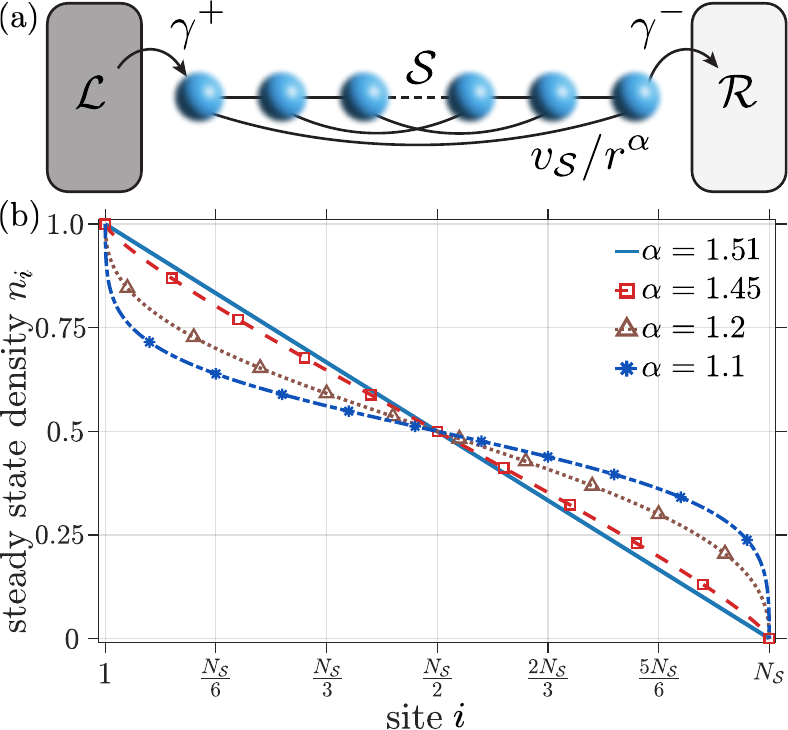}
    \caption{{\bf Boundary--drive with long--range hopping:} 
    (a) A one--dimensional lattice is held at an infinite bias with the left reservoir $\ql$ completely filled and the right reservoir $\qr$ completely empty. 
    Long--range hopping is given by a power--law, $v_\qs/r^{\alpha}$. 
    The injection and depletion rates are $\gamma^{+}$ and $\gamma^{-}$ from $\ql$ and to $\qr$, respectively. 
    Each site is subject to Markovian dephasing $\sigma$ indicated by the blurred sites. 
    (b) Stationary--state occupation, $n_i$, versus $i$ for various hopping exponents $\alpha$. 
    The solid line ($\alpha=1.51$) shows a perfectly linear density drop, while smaller $\alpha$ (dashed and dotted curves) develop pronounced curvature and boundary layers.}
    \label{fig:LongRangeSchematic}
\end{figure}

We now consider a one--dimensional lattice $\qs$ of non--interacting Fermions with long--range hopping,
\begin{equation}
    H_{\qs} =  \sum_{i=1}^{\NS-1} \sum_{r=1}^{\NS - i} \frac{v_\qs}{r^{\alpha}} \left(\cid{i} \ci{i+r} + \cid{i+r} \ci{i} \right)
\end{equation}
where $v_\qs$ is the hopping strength, $\alpha$ is the long--range exponent, $r$ is the inter--site distance (in units of the lattice spacing), and $\NS$ is the number of sites in $\qs$, see Fig.~\ref{fig:LongRangeSchematic}(a). 
The system $\qs$ has uniform onsite dephasing, i.e., $\sigma_{ij}=\sigma \, \delta_{ij}$ for $i,j \in \qs$. 
We apply an infinite bias via Markovian injection from the left  $\ql$ with rate $\gamma^{+}$, and depletion to the right $\qr$ with rate $\gamma^{-}$ at the boundaries. 

\changes{This model has been previously studied, see, e.g., Ref.~\cite{sarkar_impact_2024}, and is part of a broader class of physical systems with  long--range interactions and hopping. 
This class includes natural and artificial light--harvesting complexes~\cite{long_range_natural1, long_range_artifical1, long_range_natural2}, and models implementable with controllable couplings in experimental platforms~\cite{browaeys2020many, joshi2022observing, porras2004effective}. 
Non--local couplings can qualitatively modify equilibrium phases, ground state structure, and dynamical response~\cite{defenu2023long}. 
For instance, they can replace the ordinary diffusion under short--range models with anomalous transport, including superdiffusion and Lévy--flight behavior, as established theoretically and observed experimentally~\cite{dhawan_anomalous_2024, joshi2022observing, Knap_nonlocal}. 
They also reshape measurement--induced phase transitions (MIPT), producing new universality classes or suppressing the transition entirely~\cite{block_measurement-induced_2022, buchhold_effective_2021, Minato_MIPT_long_range}. 
Rydberg atom and trapped ion systems provide testbeds with tunable long--range couplings, with the spin--spin interactions yielding free--fermion long--range hopping as an appropriate limit~\cite{browaeys2020many}. 
Despite this progress, finite--size effects due to accessible system sizes hamper characterization: in free--fermion long--range hopping, MIPT critical points remain obscure even at $N \approx 400$~\cite{Minato_MIPT_long_range} and studies of anomalous transport are similarly constrained, impeding  the definitive determination of the super--diffusive to diffusive critical points~\cite{sarkar_impact_2024, dhawan_anomalous_2024}. 
These considerations motivate developing scalable methods and establishing benchmarks, such as the model here, for validating new algorithms.}

The model here exhibits anomalous transport, which is a stark departure from the well--known ballistic transport, where the stationary--state (SS) resistance $R_{SS} = J_{SS}^{-1}$ remains unchanged with increasing system size (the number of lattice sites $\NS$), as well as diffusive transport, where the stationary--state resistance scales as $R_{SS} \propto \NS$~\cite{dhawan_anomalous_2024,sarkar_impact_2024}. 
The anomalous transport and transition is reflected in the stationary--state density profile for different values of the long--range exponent $\alpha$, see Fig.~\ref{fig:LongRangeSchematic}(b). 
As $\alpha$ decreases from the short--range regime ($\alpha \gtrsim 1.5$) to long--range ($\alpha < 1.5$), the stationary--state equation crosses from ordinary diffusion to fractional diffusion:
\begin{eqnarray}
& & \alpha \gtrsim 1.5 \Rightarrow D\,\frac{d^2n(x)}{dx^2} = 0  \nonumber \\
& & \alpha < 1.5 \Rightarrow D_\alpha\,(-\Delta)^{\alpha/2}n(x) = 0 , \nonumber
\end{eqnarray}
where $D$ is the ordinary diffusion constant, $D_\alpha$ the anomalous diffusion constant, $(-\Delta)^{\alpha/2}$ the non--local fractional Laplacian, and $x$ the position.  
The ordinary regime yields a linear density drop, $n(x) \approx \left(1-x/L \right)$, which is approximate due to boundary effects for finite lattices.  
This regime is equivalent to dephasing and nearest‐-neighbor hopping, which follows a diffusion equation in the continuum limit~\cite{dziarmaga_non-local_2012}, with a diffusion coefficient, $D = 2 v^2/\sigma$ with $v$ the nearest--neighbor hopping and $\sigma$ the local dephasing strength. 
In the anomalous regime, the lattice is ``well mixed'' by Lévy‐flight–type jumps~\cite{schuckert_nonlocal_2020,dhar_exact_2013}, flattening the bulk and concentrating the drop into sharp boundary layers near the reservoirs---hence the non--linear density profile for $\alpha<1.5$.

The defining feature of the anomalous regime is the existence of super--diffusive transport characterized by $R_{SS} \propto \NS^{\nu}$ with $\nu<1$, and a super--diffusive--to--diffusive crossover (arguably, a phase transition) at $\alpha_c = 1.5$ in the thermodynamic limit ($\NS \rightarrow \infty$), where the diffusive regime appears for $\alpha > \alpha_c$~\cite{dhawan_anomalous_2024,sarkar_impact_2024}. 
\changes{Therefore, while the resistance diverges in both regimes, the resistivity remains finite on the diffusive side and vanishes in the thermodynamic limit on the super--diffusive side. 
The analytical estimate \(\alpha_c = 1.5\) for this change in scaling} in the thermodynamic limit follows the current operator norm, i.e., the maximum possible coherent transport~\cite{sarkar_impact_2024}.

Numerical results on $\NS \leq 1024$ indicate that $\nu$ increases with $\alpha$, saturating well above  $\alpha \gtrsim 1.6$. 
Specifically, for $1 < \alpha \leq 1.6$, the exponent scales as $\nu \approx 1.6 \, \alpha - 2$ from analyzing $\NS \leq 1024$ \cite{sarkar_impact_2024}. 
However, analytical calculations, both using the current--operator norm and mapping the corresponding Lindblad equation to fractional diffusion equation in the large dephasing limit $\sigma \gg v_{\qs}$, predict $\nu = 2 \, \alpha - 2$ in the thermodynamic limit $N_{\qs} \rightarrow \infty$~\cite{dhawan_anomalous_2024,sarkar_impact_2024}. 
This deviation of the slope of $\alpha$ dependence of $\nu$ originates from the limited accessibility of the system size, as we shall show here.
Notably, in a different setup that replaces the onsite dephasing with number--conserving B\"uttiker voltage probes, the scaling analysis obtained from a larger system size $2^{13} \lesssim \NS \lesssim 2^{14}$ achieved a similar estimate of $\alpha_c$~\cite{dhawan_anomalous_2024}.

\begin{figure}[t!]
    \centering
    \includegraphics[width=\columnwidth]{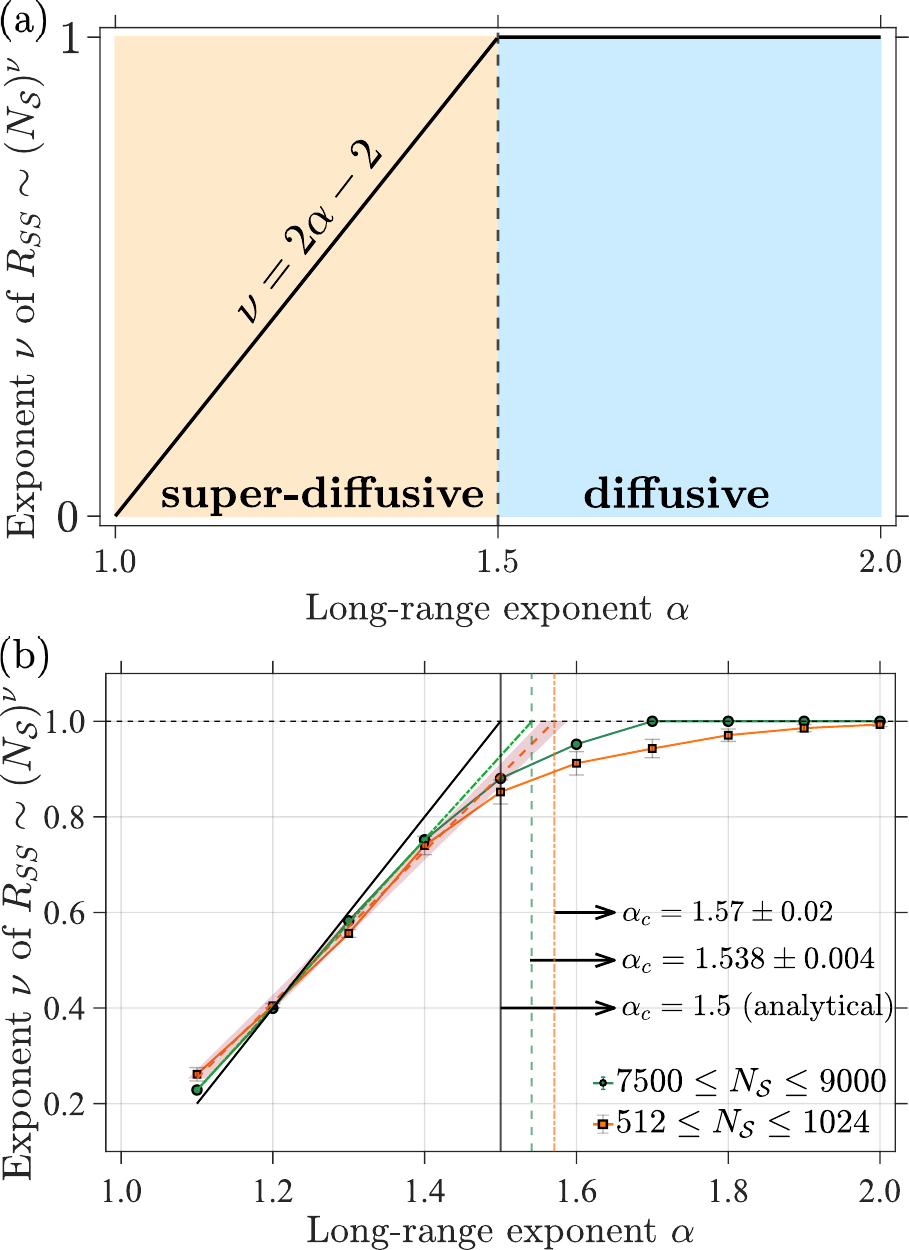}
    \caption{{\bf Super--diffusive--to--diffusive transition:} 
    (a) Phase diagram showing the transport exponent $\nu$ as a function of the long--range hopping exponent $\alpha$. 
    The stationary--state resistance scales as $R_{SS} \propto N_\qs^\nu$, where $N_\qs$ is the system size. For $\alpha \leq \alpha_c = 1.5$, the system exhibits super--diffusive transport with an exponent $\nu = 2 \, \alpha - 2$.
    For $\alpha > \alpha_c$, a transition to diffusive transport occurs, where $\nu = 1$.
    The critical value $\alpha_c =1.5$ analytically marks the boundary where the current operator norm bounds coherent transport. 
    (b) The exponent $\nu$ of the system size scaling for the resistance $R_{SS} = J_{SS}^{-1}$ in the stationary--state as a function of the long--range exponent parameter $\alpha$. 
    Orange squares (with gray error bars) and green circles (errors are smaller than the symbol) correspond to the system sizes between 512 and 1024, and between 7500 and 9000 sites, respectively. 
    The orange and green solid lines guide the eye, and the green dot--dashed and orange dashed lines correspond to the linear fit below $\alpha = 1.5$. 
    The black solid line corresponds to the analytical $\nu = 2 \, \alpha - 2$.
    Results are with $\sigma = 10^3 \, v_\qs$ and $v_\qs$ sets the frequency scale.}
    \label{fig:lng-rng-results}
\end{figure}

Using our approach for the stationary--state solution of Eq.~\eqref{eq:CMinf}, we extend the application to $\NS \approx 10^4$ for Markovian dephasing opposed to B\"uttiker probes. 
We compute $R_{SS}$ and find the scaling exponent $\nu$. 
In Fig.~\ref{fig:lng-rng-results}, we plot $\nu$, calculated from the system sizes $512 \leq \NS \leq 1024$ (orange squares) representing a small system $\Ss$ and $7500 \leq \NS \leq 9000$ (green circles) representing a large system $\Ls$, as a function of the long--range exponent $\alpha$. 
We fit the data for $\alpha < 1.5$---orange dashed line for $\Ss$ and green dots---dashed line for the larger $\Ls$. 
The former intercepts the $\nu = 1$ line---scaling exponent of $R_{SS}$ with $N_\qs$ for diffusive transport---at a critical value $\alpha_c = 1.57$ while the latter at $\alpha_c = 1.538$. 
This reduces the deviation from the analytical prediction ($\alpha_c = 1.5$) by $43~\%$, from 0.07 in $\Ss$ to 0.04 in $\Ls$. 
Furthermore, the slope of the linear fit, $\nu = \kappa \, \alpha - 2$, increases to $\kappa = 1.75 \pm 0.01$ 
for $\Ls$ from $\kappa = 1.58 \pm 0.07$ 
for $\Ss$, improving the deviation from the analytically obtained value of $\kappa = 2$ by about $40~\%$.

Moreover, the transition looks quite a bit sharper. 
For $\Ss$, the curve very slowly decays to normal diffusion for $\alpha > 1.5$, only becoming roughly $\nu=1$ at $\alpha=2$.  
Yet, for $\Ls$, the exponent more rapidly saturates to one, effectively achieving this value at $\alpha=1.7$. 
These three findings---better agreement of $\alpha_c$, better agreement of $\kappa$, and increased sharpness---provide more confidence that this is a phase transition, rather than a crossover. 

The error bar quoted above reflects the intrinsic scatter of the data about the fitted power--law form. 
Performing a least--squares fit of $\log R_{SS}$ versus $\log N_{\qs}$, the root mean square $s=\sqrt{\sum_i \epsilon_i^{\,2}/(q-k)}$ of the residuals $\epsilon_i$ is taken as a single measure of vertical dispersion, where $k$ is the number of fitted parameters (here, $k=2$). 
The value of $q$ corresponds to the number of data points between $7500 \leq N_{\qs} \leq 9000$, and the correction $(q - k)$ ensures that $s^2$ corresponds to an unbiased estimator of the variance of the underlying noise. 
This residual characterizes the typical deviation of the observations from the regression, serving as a goodness--of--fit indicator rather than a confidence interval for the slope. 
These indicators improve substantially for larger $N_\qs$. 

\section{Conclusions}
\label{sec:Conclusions}

We provided a generalized equation of motion, Eq.~\eqref{eq:MarkovianERA}, for dephasing and relaxation of fermionic networks with cross--terms for each. 
This setup covers a wide variety of physical scenarios, from extended reservoir approaches to boundary models. 
For quadratic Hamiltonians, the single--particle correlation matrix has a closed system of equations, including in the presence of dephasing and relaxation. 
We demonstrated how to solve the resulting equation efficiently. 
For an example lattice with local dephasing, long--range hopping, and boundary driving, we simulate up to $10^4$ sites. 
This enhances the accuracy of locating the critical point in anomalous quantum transport, generating better agreement with theory and further confidence that there is a phase transition. 
This approach will help improve benchmarking calculations---i.e., quadratic limits or approximations (mean--field)---for many--body simulations, such as tensor networks. 
\changes{It will also help understand the usability of ERA, such as identifying anomalous behavior, where a scaling transition in accuracy ($1/\sqrt{N}$ to $1/N$) has been observed around $10^3$ modes~\cite{wojtowicz_dual_2021}.} 
In addition, the approach is directly applicable to physical and computational networks of interest, including anomalous transport in higher dimensions and quantum networks for machine learning~\cite{lorber_using_2024}.

{\em Acknowledgments}:
We thank Yonatan Dubi and Justin Elenewski for helpful comments. 
This project was supported by the National Science Center (NCN), Poland, under projects No.~2020/38/E/ST3/00150 (S.S., G.W, M.M.R.) and~No.~2020/38/E/ST3/00269 (B.G.).
G.~W. acknowledges the Alexander von Humboldt Foundation for support under the Humboldt Research Fellowship.

\bibliography{HistoryofERA,GeneralizedDephasing}

\begin{thebibliography}{107}%
\makeatletter
\providecommand \@ifxundefined [1]{%
 \@ifx{#1\undefined}
}%
\providecommand \@ifnum [1]{%
 \ifnum #1\expandafter \@firstoftwo
 \else \expandafter \@secondoftwo
 \fi
}%
\providecommand \@ifx [1]{%
 \ifx #1\expandafter \@firstoftwo
 \else \expandafter \@secondoftwo
 \fi
}%
\providecommand \natexlab [1]{#1}%
\providecommand \enquote  [1]{``#1''}%
\providecommand \bibnamefont  [1]{#1}%
\providecommand \bibfnamefont [1]{#1}%
\providecommand \citenamefont [1]{#1}%
\providecommand \href@noop [0]{\@secondoftwo}%
\providecommand \href [0]{\begingroup \@sanitize@url \@href}%
\providecommand \@href[1]{\@@startlink{#1}\@@href}%
\providecommand \@@href[1]{\endgroup#1\@@endlink}%
\providecommand \@sanitize@url [0]{\catcode `\\12\catcode `\$12\catcode `\&12\catcode `\#12\catcode `\^12\catcode `\_12\catcode `\%12\relax}%
\providecommand \@@startlink[1]{}%
\providecommand \@@endlink[0]{}%
\providecommand \url  [0]{\begingroup\@sanitize@url \@url }%
\providecommand \@url [1]{\endgroup\@href {#1}{\urlprefix }}%
\providecommand \urlprefix  [0]{URL }%
\providecommand \Eprint [0]{\href }%
\providecommand \doibase [0]{http://dx.doi.org/}%
\providecommand \selectlanguage [0]{\@gobble}%
\providecommand \bibinfo  [0]{\@secondoftwo}%
\providecommand \bibfield  [0]{\@secondoftwo}%
\providecommand \translation [1]{[#1]}%
\providecommand \BibitemOpen [0]{}%
\providecommand \bibitemStop [0]{}%
\providecommand \bibitemNoStop [0]{.\EOS\space}%
\providecommand \EOS [0]{\spacefactor3000\relax}%
\providecommand \BibitemShut  [1]{\csname bibitem#1\endcsname}%
\let\auto@bib@innerbib\@empty
\bibitem [{\citenamefont {Subotnik}\ \emph {et~al.}(2009)\citenamefont {Subotnik}, \citenamefont {Hansen}, \citenamefont {Ratner},\ and\ \citenamefont {Nitzan}}]{subotnik_nonequilibrium_2009}%
  \BibitemOpen
  \bibfield  {author} {\bibinfo {author} {\bibfnamefont {J.~E.}\ \bibnamefont {Subotnik}}, \bibinfo {author} {\bibfnamefont {T.}~\bibnamefont {Hansen}}, \bibinfo {author} {\bibfnamefont {M.~A.}\ \bibnamefont {Ratner}}, \ and\ \bibinfo {author} {\bibfnamefont {A.}~\bibnamefont {Nitzan}},\ }\href {http://aip.scitation.org/doi/10.1063/1.3109898} {\bibfield  {journal} {\bibinfo  {journal} {J. Chem. Phys.}\ }\textbf {\bibinfo {volume} {130}},\ \bibinfo {pages} {144105} (\bibinfo {year} {2009})}\BibitemShut {NoStop}%
\bibitem [{\citenamefont {S\'anchez}\ \emph {et~al.}(2006)\citenamefont {S\'anchez}, \citenamefont {Stamenova}, \citenamefont {Sanvito}, \citenamefont {Bowler}, \citenamefont {Horsfield},\ and\ \citenamefont {Todorov}}]{sanchez_molecular_2006}%
  \BibitemOpen
  \bibfield  {author} {\bibinfo {author} {\bibfnamefont {C.~G.}\ \bibnamefont {S\'anchez}}, \bibinfo {author} {\bibfnamefont {M.}~\bibnamefont {Stamenova}}, \bibinfo {author} {\bibfnamefont {S.}~\bibnamefont {Sanvito}}, \bibinfo {author} {\bibfnamefont {D.~R.}\ \bibnamefont {Bowler}}, \bibinfo {author} {\bibfnamefont {A.~P.}\ \bibnamefont {Horsfield}}, \ and\ \bibinfo {author} {\bibfnamefont {T.~N.}\ \bibnamefont {Todorov}},\ }\href {https://aip.scitation.org/doi/10.1063/1.2202329} {\bibfield  {journal} {\bibinfo  {journal} {J. Chem. Phys.}\ }\textbf {\bibinfo {volume} {124}},\ \bibinfo {pages} {214708} (\bibinfo {year} {2006})}\BibitemShut {NoStop}%
\bibitem [{\citenamefont {McEniry}\ \emph {et~al.}(2007)\citenamefont {McEniry}, \citenamefont {Bowler}, \citenamefont {Dundas}, \citenamefont {Horsfield}, \citenamefont {Sánchez},\ and\ \citenamefont {Todorov}}]{mceniry_dynamical_2007}%
  \BibitemOpen
  \bibfield  {author} {\bibinfo {author} {\bibfnamefont {E.~J.}\ \bibnamefont {McEniry}}, \bibinfo {author} {\bibfnamefont {D.~R.}\ \bibnamefont {Bowler}}, \bibinfo {author} {\bibfnamefont {D.}~\bibnamefont {Dundas}}, \bibinfo {author} {\bibfnamefont {A.~P.}\ \bibnamefont {Horsfield}}, \bibinfo {author} {\bibfnamefont {C.~G.}\ \bibnamefont {Sánchez}}, \ and\ \bibinfo {author} {\bibfnamefont {T.~N.}\ \bibnamefont {Todorov}},\ }\href {https://dx.doi.org/10.1088/0953-8984/19/19/196201} {\bibfield  {journal} {\bibinfo  {journal} {J. Phys.: Condens. Matter}\ }\textbf {\bibinfo {volume} {19}},\ \bibinfo {pages} {196201} (\bibinfo {year} {2007})}\BibitemShut {NoStop}%
\bibitem [{\citenamefont {Cazalilla}\ and\ \citenamefont {Marston}(2002)}]{cazalilla_time-dependent_2002}%
  \BibitemOpen
  \bibfield  {author} {\bibinfo {author} {\bibfnamefont {M.~A.}\ \bibnamefont {Cazalilla}}\ and\ \bibinfo {author} {\bibfnamefont {J.~B.}\ \bibnamefont {Marston}},\ }\href {https://link.aps.org/doi/10.1103/PhysRevLett.88.256403} {\bibfield  {journal} {\bibinfo  {journal} {Phys. Rev. Lett.}\ }\textbf {\bibinfo {volume} {88}},\ \bibinfo {pages} {256403} (\bibinfo {year} {2002})}\BibitemShut {NoStop}%
\bibitem [{\citenamefont {Zwolak}\ and\ \citenamefont {Vidal}(2004)}]{zwolak_mixed-state_2004}%
  \BibitemOpen
  \bibfield  {author} {\bibinfo {author} {\bibfnamefont {M.}~\bibnamefont {Zwolak}}\ and\ \bibinfo {author} {\bibfnamefont {G.}~\bibnamefont {Vidal}},\ }\href {https://link.aps.org/doi/10.1103/PhysRevLett.93.207205} {\bibfield  {journal} {\bibinfo  {journal} {Phys. Rev. Lett.}\ }\textbf {\bibinfo {volume} {93}},\ \bibinfo {pages} {207205} (\bibinfo {year} {2004})}\BibitemShut {NoStop}%
\bibitem [{\citenamefont {Ventra}\ and\ \citenamefont {Todorov}(2004)}]{ventra_transport_2004}%
  \BibitemOpen
  \bibfield  {author} {\bibinfo {author} {\bibfnamefont {M.~D.}\ \bibnamefont {Ventra}}\ and\ \bibinfo {author} {\bibfnamefont {T.~N.}\ \bibnamefont {Todorov}},\ }\href {https://doi.org/10.1088%2F0953-8984%2F16%2F45%2F024} {\bibfield  {journal} {\bibinfo  {journal} {J. Phys.: Condens. Matter}\ }\textbf {\bibinfo {volume} {16}},\ \bibinfo {pages} {8025} (\bibinfo {year} {2004})}\BibitemShut {NoStop}%
\bibitem [{\citenamefont {Bohr}\ \emph {et~al.}(2005)\citenamefont {Bohr}, \citenamefont {Schmitteckert},\ and\ \citenamefont {W\"olfle}}]{bohr_dmrg_2005}%
  \BibitemOpen
  \bibfield  {author} {\bibinfo {author} {\bibfnamefont {D.}~\bibnamefont {Bohr}}, \bibinfo {author} {\bibfnamefont {P.}~\bibnamefont {Schmitteckert}}, \ and\ \bibinfo {author} {\bibfnamefont {P.}~\bibnamefont {W\"olfle}},\ }\href {https://iopscience.iop.org/article/10.1209/epl/i2005-10377-6/meta} {\bibfield  {journal} {\bibinfo  {journal} {EPL (Europhysics Letters)}\ }\textbf {\bibinfo {volume} {73}},\ \bibinfo {pages} {246} (\bibinfo {year} {2005})}\BibitemShut {NoStop}%
\bibitem [{\citenamefont {Bushong}\ \emph {et~al.}(2005)\citenamefont {Bushong}, \citenamefont {Sai},\ and\ \citenamefont {Di~Ventra}}]{bushong_approach_2005}%
  \BibitemOpen
  \bibfield  {author} {\bibinfo {author} {\bibfnamefont {N.}~\bibnamefont {Bushong}}, \bibinfo {author} {\bibfnamefont {N.}~\bibnamefont {Sai}}, \ and\ \bibinfo {author} {\bibfnamefont {M.}~\bibnamefont {Di~Ventra}},\ }\href {https://pubs.acs.org/doi/10.1021/nl0520157} {\bibfield  {journal} {\bibinfo  {journal} {Nano Lett.}\ }\textbf {\bibinfo {volume} {5}},\ \bibinfo {pages} {2569} (\bibinfo {year} {2005})}\BibitemShut {NoStop}%
\bibitem [{\citenamefont {Gobert}\ \emph {et~al.}(2005)\citenamefont {Gobert}, \citenamefont {Kollath}, \citenamefont {Schollw\"ock},\ and\ \citenamefont {Schütz}}]{gobert_real-time_2005}%
  \BibitemOpen
  \bibfield  {author} {\bibinfo {author} {\bibfnamefont {D.}~\bibnamefont {Gobert}}, \bibinfo {author} {\bibfnamefont {C.}~\bibnamefont {Kollath}}, \bibinfo {author} {\bibfnamefont {U.}~\bibnamefont {Schollw\"ock}}, \ and\ \bibinfo {author} {\bibfnamefont {G.}~\bibnamefont {Schütz}},\ }\href {https://link.aps.org/doi/10.1103/PhysRevE.71.036102} {\bibfield  {journal} {\bibinfo  {journal} {Phys. Rev. E}\ }\textbf {\bibinfo {volume} {71}},\ \bibinfo {pages} {036102} (\bibinfo {year} {2005})}\BibitemShut {NoStop}%
\bibitem [{\citenamefont {Al-Hassanieh}\ \emph {et~al.}(2006)\citenamefont {Al-Hassanieh}, \citenamefont {Feiguin}, \citenamefont {Riera}, \citenamefont {Büsser},\ and\ \citenamefont {Dagotto}}]{al-hassanieh_adaptive_2006}%
  \BibitemOpen
  \bibfield  {author} {\bibinfo {author} {\bibfnamefont {K.~A.}\ \bibnamefont {Al-Hassanieh}}, \bibinfo {author} {\bibfnamefont {A.~E.}\ \bibnamefont {Feiguin}}, \bibinfo {author} {\bibfnamefont {J.~A.}\ \bibnamefont {Riera}}, \bibinfo {author} {\bibfnamefont {C.~A.}\ \bibnamefont {Büsser}}, \ and\ \bibinfo {author} {\bibfnamefont {E.}~\bibnamefont {Dagotto}},\ }\href {https://link.aps.org/doi/10.1103/PhysRevB.73.195304} {\bibfield  {journal} {\bibinfo  {journal} {Phys. Rev. B}\ }\textbf {\bibinfo {volume} {73}},\ \bibinfo {pages} {195304} (\bibinfo {year} {2006})}\BibitemShut {NoStop}%
\bibitem [{\citenamefont {Cheng}\ \emph {et~al.}(2006)\citenamefont {Cheng}, \citenamefont {Evans},\ and\ \citenamefont {Van~Voorhis}}]{cheng_simulating_2006}%
  \BibitemOpen
  \bibfield  {author} {\bibinfo {author} {\bibfnamefont {C.-L.}\ \bibnamefont {Cheng}}, \bibinfo {author} {\bibfnamefont {J.~S.}\ \bibnamefont {Evans}}, \ and\ \bibinfo {author} {\bibfnamefont {T.}~\bibnamefont {Van~Voorhis}},\ }\href {https://link.aps.org/doi/10.1103/PhysRevB.74.155112} {\bibfield  {journal} {\bibinfo  {journal} {Phys. Rev. B}\ }\textbf {\bibinfo {volume} {74}},\ \bibinfo {pages} {155112} (\bibinfo {year} {2006})}\BibitemShut {NoStop}%
\bibitem [{\citenamefont {Schneider}\ and\ \citenamefont {Schmitteckert}(2006)}]{schneider_conductance_2006}%
  \BibitemOpen
  \bibfield  {author} {\bibinfo {author} {\bibfnamefont {G.}~\bibnamefont {Schneider}}\ and\ \bibinfo {author} {\bibfnamefont {P.}~\bibnamefont {Schmitteckert}},\ }\href {http://arxiv.org/abs/cond-mat/0601389} {\bibfield  {journal} {\bibinfo  {journal} {arXiv:cond-mat/0601389}\ } (\bibinfo {year} {2006})}\BibitemShut {NoStop}%
\bibitem [{\citenamefont {Schmitteckert}\ and\ \citenamefont {Schneider}(2006)}]{schmitteckert_signal_2006}%
  \BibitemOpen
  \bibfield  {author} {\bibinfo {author} {\bibfnamefont {P.}~\bibnamefont {Schmitteckert}}\ and\ \bibinfo {author} {\bibfnamefont {G.}~\bibnamefont {Schneider}},\ }in\ \href@noop {} {\emph {\bibinfo {booktitle} {High {Performance} {Computing} in {Science} and {Engineering}}}},\ \bibinfo {editor} {edited by\ \bibinfo {editor} {\bibfnamefont {W.~E.}\ \bibnamefont {Nagel}}, \bibinfo {editor} {\bibfnamefont {W.}~\bibnamefont {J\"ager}}, \ and\ \bibinfo {editor} {\bibfnamefont {M.}~\bibnamefont {Resch}}}\ (\bibinfo  {publisher} {Springer, Berlin},\ \bibinfo {year} {2006})\ pp.\ \bibinfo {pages} {113--126}\BibitemShut {NoStop}%
\bibitem [{\citenamefont {Sai}\ \emph {et~al.}(2007)\citenamefont {Sai}, \citenamefont {Bushong}, \citenamefont {Hatcher},\ and\ \citenamefont {Di~Ventra}}]{sai_microscopic_2007}%
  \BibitemOpen
  \bibfield  {author} {\bibinfo {author} {\bibfnamefont {N.}~\bibnamefont {Sai}}, \bibinfo {author} {\bibfnamefont {N.}~\bibnamefont {Bushong}}, \bibinfo {author} {\bibfnamefont {R.}~\bibnamefont {Hatcher}}, \ and\ \bibinfo {author} {\bibfnamefont {M.}~\bibnamefont {Di~Ventra}},\ }\href {https://link.aps.org/doi/10.1103/PhysRevB.75.115410} {\bibfield  {journal} {\bibinfo  {journal} {Phys. Rev. B}\ }\textbf {\bibinfo {volume} {75}},\ \bibinfo {pages} {115410} (\bibinfo {year} {2007})}\BibitemShut {NoStop}%
\bibitem [{\citenamefont {Bohr}\ and\ \citenamefont {Schmitteckert}(2007)}]{bohr_strong_2007}%
  \BibitemOpen
  \bibfield  {author} {\bibinfo {author} {\bibfnamefont {D.}~\bibnamefont {Bohr}}\ and\ \bibinfo {author} {\bibfnamefont {P.}~\bibnamefont {Schmitteckert}},\ }\href {https://link.aps.org/doi/10.1103/PhysRevB.75.241103} {\bibfield  {journal} {\bibinfo  {journal} {Phys. Rev. B}\ }\textbf {\bibinfo {volume} {75}},\ \bibinfo {pages} {241103} (\bibinfo {year} {2007})}\BibitemShut {NoStop}%
\bibitem [{\citenamefont {Zwolak}(2008{\natexlab{a}})}]{zwolak_finite_2008}%
  \BibitemOpen
  \bibfield  {author} {\bibinfo {author} {\bibfnamefont {M.}~\bibnamefont {Zwolak}},\ }\href {https://aip.scitation.org/doi/full/10.1063/1.2976008} {\bibfield  {journal} {\bibinfo  {journal} {J. Chem. Phys.}\ }\textbf {\bibinfo {volume} {129}},\ \bibinfo {pages} {101101} (\bibinfo {year} {2008}{\natexlab{a}})}\BibitemShut {NoStop}%
\bibitem [{\citenamefont {Dias~da Silva}\ \emph {et~al.}(2008)\citenamefont {Dias~da Silva}, \citenamefont {Heidrich-Meisner}, \citenamefont {Feiguin}, \citenamefont {Büsser}, \citenamefont {Martins}, \citenamefont {Anda},\ and\ \citenamefont {Dagotto}}]{dias_da_silva_transport_2008}%
  \BibitemOpen
  \bibfield  {author} {\bibinfo {author} {\bibfnamefont {L.~G. G.~V.}\ \bibnamefont {Dias~da Silva}}, \bibinfo {author} {\bibfnamefont {F.}~\bibnamefont {Heidrich-Meisner}}, \bibinfo {author} {\bibfnamefont {A.~E.}\ \bibnamefont {Feiguin}}, \bibinfo {author} {\bibfnamefont {C.~A.}\ \bibnamefont {Büsser}}, \bibinfo {author} {\bibfnamefont {G.~B.}\ \bibnamefont {Martins}}, \bibinfo {author} {\bibfnamefont {E.~V.}\ \bibnamefont {Anda}}, \ and\ \bibinfo {author} {\bibfnamefont {E.}~\bibnamefont {Dagotto}},\ }\href {https://link.aps.org/doi/10.1103/PhysRevB.78.195317} {\bibfield  {journal} {\bibinfo  {journal} {Phys. Rev. B}\ }\textbf {\bibinfo {volume} {78}},\ \bibinfo {pages} {195317} (\bibinfo {year} {2008})}\BibitemShut {NoStop}%
\bibitem [{\citenamefont {Heidrich-Meisner}\ \emph {et~al.}(2009)\citenamefont {Heidrich-Meisner}, \citenamefont {Feiguin},\ and\ \citenamefont {Dagotto}}]{heidrich-meisner_real-time_2009}%
  \BibitemOpen
  \bibfield  {author} {\bibinfo {author} {\bibfnamefont {F.}~\bibnamefont {Heidrich-Meisner}}, \bibinfo {author} {\bibfnamefont {A.~E.}\ \bibnamefont {Feiguin}}, \ and\ \bibinfo {author} {\bibfnamefont {E.}~\bibnamefont {Dagotto}},\ }\href {https://link.aps.org/doi/10.1103/PhysRevB.79.235336} {\bibfield  {journal} {\bibinfo  {journal} {Phys. Rev. B}\ }\textbf {\bibinfo {volume} {79}},\ \bibinfo {pages} {235336} (\bibinfo {year} {2009})}\BibitemShut {NoStop}%
\bibitem [{\citenamefont {Bransch\"adel}\ \emph {et~al.}(2010)\citenamefont {Bransch\"adel}, \citenamefont {Schneider},\ and\ \citenamefont {Schmitteckert}}]{branschadel_conductance_2010}%
  \BibitemOpen
  \bibfield  {author} {\bibinfo {author} {\bibfnamefont {A.}~\bibnamefont {Bransch\"adel}}, \bibinfo {author} {\bibfnamefont {G.}~\bibnamefont {Schneider}}, \ and\ \bibinfo {author} {\bibfnamefont {P.}~\bibnamefont {Schmitteckert}},\ }\href {http://dx.doi.org/10.1002/andp.201000017} {\bibfield  {journal} {\bibinfo  {journal} {Ann. Phys. (Berlin)}\ }\textbf {\bibinfo {volume} {522}},\ \bibinfo {pages} {657} (\bibinfo {year} {2010})}\BibitemShut {NoStop}%
\bibitem [{\citenamefont {Chien}\ \emph {et~al.}(2012)\citenamefont {Chien}, \citenamefont {Zwolak},\ and\ \citenamefont {Di~Ventra}}]{chien_bosonic_2012}%
  \BibitemOpen
  \bibfield  {author} {\bibinfo {author} {\bibfnamefont {C.-C.}\ \bibnamefont {Chien}}, \bibinfo {author} {\bibfnamefont {M.}~\bibnamefont {Zwolak}}, \ and\ \bibinfo {author} {\bibfnamefont {M.}~\bibnamefont {Di~Ventra}},\ }\href {https://link.aps.org/doi/10.1103/PhysRevA.85.041601} {\bibfield  {journal} {\bibinfo  {journal} {Phys. Rev. A}\ }\textbf {\bibinfo {volume} {85}},\ \bibinfo {pages} {041601} (\bibinfo {year} {2012})}\BibitemShut {NoStop}%
\bibitem [{\citenamefont {Chien}\ \emph {et~al.}(2013{\natexlab{a}})\citenamefont {Chien}, \citenamefont {Gruss}, \citenamefont {Ventra},\ and\ \citenamefont {Zwolak}}]{chien_interaction-induced_2013}%
  \BibitemOpen
  \bibfield  {author} {\bibinfo {author} {\bibfnamefont {C.-C.}\ \bibnamefont {Chien}}, \bibinfo {author} {\bibfnamefont {D.}~\bibnamefont {Gruss}}, \bibinfo {author} {\bibfnamefont {M.~D.}\ \bibnamefont {Ventra}}, \ and\ \bibinfo {author} {\bibfnamefont {M.}~\bibnamefont {Zwolak}},\ }\href {https://iopscience.iop.org/article/10.1088/1367-2630/15/6/063026} {\bibfield  {journal} {\bibinfo  {journal} {New J. Phys.}\ }\textbf {\bibinfo {volume} {15}},\ \bibinfo {pages} {063026} (\bibinfo {year} {2013}{\natexlab{a}})}\BibitemShut {NoStop}%
\bibitem [{\citenamefont {Chien}\ \emph {et~al.}(2014)\citenamefont {Chien}, \citenamefont {Di~Ventra},\ and\ \citenamefont {Zwolak}}]{chien_landauer_2014}%
  \BibitemOpen
  \bibfield  {author} {\bibinfo {author} {\bibfnamefont {C.-C.}\ \bibnamefont {Chien}}, \bibinfo {author} {\bibfnamefont {M.}~\bibnamefont {Di~Ventra}}, \ and\ \bibinfo {author} {\bibfnamefont {M.}~\bibnamefont {Zwolak}},\ }\href {https://link.aps.org/doi/10.1103/PhysRevA.90.023624} {\bibfield  {journal} {\bibinfo  {journal} {Phys. Rev. A}\ }\textbf {\bibinfo {volume} {90}},\ \bibinfo {pages} {023624} (\bibinfo {year} {2014})}\BibitemShut {NoStop}%
\bibitem [{\citenamefont {Zwolak}(2018)}]{zwolak_communication_2018}%
  \BibitemOpen
  \bibfield  {author} {\bibinfo {author} {\bibfnamefont {M.}~\bibnamefont {Zwolak}},\ }\href {https://aip.scitation.org/doi/10.1063/1.5061759} {\bibfield  {journal} {\bibinfo  {journal} {J. Chem. Phys.}\ }\textbf {\bibinfo {volume} {149}},\ \bibinfo {pages} {241102} (\bibinfo {year} {2018})}\BibitemShut {NoStop}%
\bibitem [{\citenamefont {Gruss}\ \emph {et~al.}(2018{\natexlab{a}})\citenamefont {Gruss}, \citenamefont {Chien}, \citenamefont {Barreiro}, \citenamefont {Di~Ventra},\ and\ \citenamefont {Zwolak}}]{gruss_energy-resolved_2018}%
  \BibitemOpen
  \bibfield  {author} {\bibinfo {author} {\bibfnamefont {D.}~\bibnamefont {Gruss}}, \bibinfo {author} {\bibfnamefont {C.-C.}\ \bibnamefont {Chien}}, \bibinfo {author} {\bibfnamefont {J.~T.}\ \bibnamefont {Barreiro}}, \bibinfo {author} {\bibfnamefont {M.}~\bibnamefont {Di~Ventra}}, \ and\ \bibinfo {author} {\bibfnamefont {M.}~\bibnamefont {Zwolak}},\ }\href {https://dx.doi.org/10.1088/1367-2630/aaedcf} {\bibfield  {journal} {\bibinfo  {journal} {New J. Phys.}\ }\textbf {\bibinfo {volume} {20}},\ \bibinfo {pages} {115005} (\bibinfo {year} {2018}{\natexlab{a}})}\BibitemShut {NoStop}%
\bibitem [{\citenamefont {Zelovich}\ \emph {et~al.}(2014)\citenamefont {Zelovich}, \citenamefont {Kronik},\ and\ \citenamefont {Hod}}]{zelovich_state_2014}%
  \BibitemOpen
  \bibfield  {author} {\bibinfo {author} {\bibfnamefont {T.}~\bibnamefont {Zelovich}}, \bibinfo {author} {\bibfnamefont {L.}~\bibnamefont {Kronik}}, \ and\ \bibinfo {author} {\bibfnamefont {O.}~\bibnamefont {Hod}},\ }\href {https://doi.org/10.1021/ct500135e} {\bibfield  {journal} {\bibinfo  {journal} {J. Chem. Theory Comput.}\ }\textbf {\bibinfo {volume} {10}},\ \bibinfo {pages} {2927} (\bibinfo {year} {2014})}\BibitemShut {NoStop}%
\bibitem [{\citenamefont {Zelovich}\ \emph {et~al.}(2015)\citenamefont {Zelovich}, \citenamefont {Kronik},\ and\ \citenamefont {Hod}}]{zelovich_moleculelead_2015}%
  \BibitemOpen
  \bibfield  {author} {\bibinfo {author} {\bibfnamefont {T.}~\bibnamefont {Zelovich}}, \bibinfo {author} {\bibfnamefont {L.}~\bibnamefont {Kronik}}, \ and\ \bibinfo {author} {\bibfnamefont {O.}~\bibnamefont {Hod}},\ }\href {http://dx.doi.org/10.1021/acs.jctc.5b00612} {\bibfield  {journal} {\bibinfo  {journal} {J. Chem. Theory Comput.}\ }\textbf {\bibinfo {volume} {11}},\ \bibinfo {pages} {4861} (\bibinfo {year} {2015})}\BibitemShut {NoStop}%
\bibitem [{\citenamefont {Zelovich}\ \emph {et~al.}(2016)\citenamefont {Zelovich}, \citenamefont {Kronik},\ and\ \citenamefont {Hod}}]{zelovich_driven_2016}%
  \BibitemOpen
  \bibfield  {author} {\bibinfo {author} {\bibfnamefont {T.}~\bibnamefont {Zelovich}}, \bibinfo {author} {\bibfnamefont {L.}~\bibnamefont {Kronik}}, \ and\ \bibinfo {author} {\bibfnamefont {O.}~\bibnamefont {Hod}},\ }\href {http://dx.doi.org/10.1021/acs.jpcc.6b03838} {\bibfield  {journal} {\bibinfo  {journal} {J. Phys. Chem. C}\ }\textbf {\bibinfo {volume} {120}},\ \bibinfo {pages} {15052} (\bibinfo {year} {2016})}\BibitemShut {NoStop}%
\bibitem [{\citenamefont {Hod}\ \emph {et~al.}(2016)\citenamefont {Hod}, \citenamefont {Rodríguez-Rosario}, \citenamefont {Zelovich},\ and\ \citenamefont {Frauenheim}}]{hod_driven_2016}%
  \BibitemOpen
  \bibfield  {author} {\bibinfo {author} {\bibfnamefont {O.}~\bibnamefont {Hod}}, \bibinfo {author} {\bibfnamefont {C.~A.}\ \bibnamefont {Rodríguez-Rosario}}, \bibinfo {author} {\bibfnamefont {T.}~\bibnamefont {Zelovich}}, \ and\ \bibinfo {author} {\bibfnamefont {T.}~\bibnamefont {Frauenheim}},\ }\href {https://doi.org/10.1021/acs.jpca.5b12212} {\bibfield  {journal} {\bibinfo  {journal} {J. Phys. Chem. A}\ }\textbf {\bibinfo {volume} {120}},\ \bibinfo {pages} {3278} (\bibinfo {year} {2016})}\BibitemShut {NoStop}%
\bibitem [{\citenamefont {Zelovich}\ \emph {et~al.}(2017)\citenamefont {Zelovich}, \citenamefont {Hansen}, \citenamefont {Liu}, \citenamefont {Neaton}, \citenamefont {Kronik},\ and\ \citenamefont {Hod}}]{zelovich_parameter-free_2017}%
  \BibitemOpen
  \bibfield  {author} {\bibinfo {author} {\bibfnamefont {T.}~\bibnamefont {Zelovich}}, \bibinfo {author} {\bibfnamefont {T.}~\bibnamefont {Hansen}}, \bibinfo {author} {\bibfnamefont {Z.-F.}\ \bibnamefont {Liu}}, \bibinfo {author} {\bibfnamefont {J.~B.}\ \bibnamefont {Neaton}}, \bibinfo {author} {\bibfnamefont {L.}~\bibnamefont {Kronik}}, \ and\ \bibinfo {author} {\bibfnamefont {O.}~\bibnamefont {Hod}},\ }\href {https://doi.org/10.1063/1.4976731} {\bibfield  {journal} {\bibinfo  {journal} {J. Chem. Phys.}\ }\textbf {\bibinfo {volume} {146}},\ \bibinfo {pages} {092331} (\bibinfo {year} {2017})}\BibitemShut {NoStop}%
\bibitem [{\citenamefont {Hod}\ and\ \citenamefont {Kronik}(2023)}]{hod_driven_2023}%
  \BibitemOpen
  \bibfield  {author} {\bibinfo {author} {\bibfnamefont {O.}~\bibnamefont {Hod}}\ and\ \bibinfo {author} {\bibfnamefont {L.}~\bibnamefont {Kronik}},\ }\href {https://onlinelibrary.wiley.com/doi/abs/10.1002/ijch.202300058} {\bibfield  {journal} {\bibinfo  {journal} {Isr. J. Chem.}\ }\textbf {\bibinfo {volume} {63}},\ \bibinfo {pages} {e202300058} (\bibinfo {year} {2023})}\BibitemShut {NoStop}%
\bibitem [{\citenamefont {Gruss}\ \emph {et~al.}(2016)\citenamefont {Gruss}, \citenamefont {Velizhanin},\ and\ \citenamefont {Zwolak}}]{gruss_landauers_2016}%
  \BibitemOpen
  \bibfield  {author} {\bibinfo {author} {\bibfnamefont {D.}~\bibnamefont {Gruss}}, \bibinfo {author} {\bibfnamefont {K.~A.}\ \bibnamefont {Velizhanin}}, \ and\ \bibinfo {author} {\bibfnamefont {M.}~\bibnamefont {Zwolak}},\ }\href {http://www.nature.com/srep/2016/160420/srep24514/full/srep24514.html} {\bibfield  {journal} {\bibinfo  {journal} {Sci. Rep.}\ }\textbf {\bibinfo {volume} {6}},\ \bibinfo {pages} {24514} (\bibinfo {year} {2016})}\BibitemShut {NoStop}%
\bibitem [{\citenamefont {Gruss}\ \emph {et~al.}(2017)\citenamefont {Gruss}, \citenamefont {Smolyanitsky},\ and\ \citenamefont {Zwolak}}]{gruss_communication_2017}%
  \BibitemOpen
  \bibfield  {author} {\bibinfo {author} {\bibfnamefont {D.}~\bibnamefont {Gruss}}, \bibinfo {author} {\bibfnamefont {A.}~\bibnamefont {Smolyanitsky}}, \ and\ \bibinfo {author} {\bibfnamefont {M.}~\bibnamefont {Zwolak}},\ }\href {https://aip.scitation.org/doi/abs/10.1063/1.4997022} {\bibfield  {journal} {\bibinfo  {journal} {J. Chem. Phys.}\ }\textbf {\bibinfo {volume} {147}},\ \bibinfo {pages} {141102} (\bibinfo {year} {2017})}\BibitemShut {NoStop}%
\bibitem [{\citenamefont {Elenewski}\ \emph {et~al.}(2017)\citenamefont {Elenewski}, \citenamefont {Gruss},\ and\ \citenamefont {Zwolak}}]{elenewski_communication_2017}%
  \BibitemOpen
  \bibfield  {author} {\bibinfo {author} {\bibfnamefont {J.~E.}\ \bibnamefont {Elenewski}}, \bibinfo {author} {\bibfnamefont {D.}~\bibnamefont {Gruss}}, \ and\ \bibinfo {author} {\bibfnamefont {M.}~\bibnamefont {Zwolak}},\ }\href {http://aip.scitation.org/doi/10.1063/1.5000747} {\bibfield  {journal} {\bibinfo  {journal} {J. Chem. Phys.}\ }\textbf {\bibinfo {volume} {147}},\ \bibinfo {pages} {151101} (\bibinfo {year} {2017})}\BibitemShut {NoStop}%
\bibitem [{\citenamefont {Gruss}\ \emph {et~al.}(2018{\natexlab{b}})\citenamefont {Gruss}, \citenamefont {Smolyanitsky},\ and\ \citenamefont {Zwolak}}]{gruss_graphene_2018}%
  \BibitemOpen
  \bibfield  {author} {\bibinfo {author} {\bibfnamefont {D.}~\bibnamefont {Gruss}}, \bibinfo {author} {\bibfnamefont {A.}~\bibnamefont {Smolyanitsky}}, \ and\ \bibinfo {author} {\bibfnamefont {M.}~\bibnamefont {Zwolak}},\ }\href {http://arxiv.org/abs/1804.02701} {\bibfield  {journal} {\bibinfo  {journal} {arXiv:1804.02701}\ } (\bibinfo {year} {2018}{\natexlab{b}})}\BibitemShut {NoStop}%
\bibitem [{\citenamefont {Zwolak}(2020{\natexlab{a}})}]{zwolak_analytic_2020}%
  \BibitemOpen
  \bibfield  {author} {\bibinfo {author} {\bibfnamefont {M.}~\bibnamefont {Zwolak}},\ }\href {http://aip.scitation.org/doi/10.1063/5.0029223} {\bibfield  {journal} {\bibinfo  {journal} {J. Chem. Phys.}\ }\textbf {\bibinfo {volume} {153}},\ \bibinfo {pages} {224107} (\bibinfo {year} {2020}{\natexlab{a}})}\BibitemShut {NoStop}%
\bibitem [{\citenamefont {Zwolak}(2020{\natexlab{b}})}]{zwolak_comment_2020}%
  \BibitemOpen
  \bibfield  {author} {\bibinfo {author} {\bibfnamefont {M.}~\bibnamefont {Zwolak}},\ }\href {http://arxiv.org/abs/2009.04466} {\bibfield  {journal} {\bibinfo  {journal} {arXiv:2009.04466}\ } (\bibinfo {year} {2020}{\natexlab{b}})}\BibitemShut {NoStop}%
\bibitem [{\citenamefont {Elenewski}\ \emph {et~al.}(2021)\citenamefont {Elenewski}, \citenamefont {W\'{o}jtowicz}, \citenamefont {Rams},\ and\ \citenamefont {Zwolak}}]{elenewski_performance_2021}%
  \BibitemOpen
  \bibfield  {author} {\bibinfo {author} {\bibfnamefont {J.~E.}\ \bibnamefont {Elenewski}}, \bibinfo {author} {\bibfnamefont {G.}~\bibnamefont {W\'{o}jtowicz}}, \bibinfo {author} {\bibfnamefont {M.~M.}\ \bibnamefont {Rams}}, \ and\ \bibinfo {author} {\bibfnamefont {M.}~\bibnamefont {Zwolak}},\ }\href {https://aip.scitation.org/doi/10.1063/5.0065799} {\bibfield  {journal} {\bibinfo  {journal} {J. Chem. Phys.}\ }\textbf {\bibinfo {volume} {155}},\ \bibinfo {pages} {124117} (\bibinfo {year} {2021})}\BibitemShut {NoStop}%
\bibitem [{\citenamefont {Wójtowicz}\ \emph {et~al.}(2021)\citenamefont {Wójtowicz}, \citenamefont {Elenewski}, \citenamefont {Rams},\ and\ \citenamefont {Zwolak}}]{wojtowicz_dual_2021}%
  \BibitemOpen
  \bibfield  {author} {\bibinfo {author} {\bibfnamefont {G.}~\bibnamefont {Wójtowicz}}, \bibinfo {author} {\bibfnamefont {J.~E.}\ \bibnamefont {Elenewski}}, \bibinfo {author} {\bibfnamefont {M.~M.}\ \bibnamefont {Rams}}, \ and\ \bibinfo {author} {\bibfnamefont {M.}~\bibnamefont {Zwolak}},\ }\href {https://link.aps.org/doi/10.1103/PhysRevB.104.165131} {\bibfield  {journal} {\bibinfo  {journal} {Phys. Rev. B}\ }\textbf {\bibinfo {volume} {104}},\ \bibinfo {pages} {165131} (\bibinfo {year} {2021})}\BibitemShut {NoStop}%
\bibitem [{\citenamefont {Wójtowicz}\ \emph {et~al.}(2023)\citenamefont {Wójtowicz}, \citenamefont {Purkayastha}, \citenamefont {Zwolak},\ and\ \citenamefont {Rams}}]{wojtowicz_accumulative_2023}%
  \BibitemOpen
  \bibfield  {author} {\bibinfo {author} {\bibfnamefont {G.}~\bibnamefont {Wójtowicz}}, \bibinfo {author} {\bibfnamefont {A.}~\bibnamefont {Purkayastha}}, \bibinfo {author} {\bibfnamefont {M.}~\bibnamefont {Zwolak}}, \ and\ \bibinfo {author} {\bibfnamefont {M.~M.}\ \bibnamefont {Rams}},\ }\href {https://link.aps.org/doi/10.1103/PhysRevB.107.035150} {\bibfield  {journal} {\bibinfo  {journal} {Phys. Rev. B}\ }\textbf {\bibinfo {volume} {107}},\ \bibinfo {pages} {035150} (\bibinfo {year} {2023})}\BibitemShut {NoStop}%
\bibitem [{\citenamefont {Dzhioev}\ and\ \citenamefont {Kosov}(2011)}]{dzhioev_super-fermion_2011}%
  \BibitemOpen
  \bibfield  {author} {\bibinfo {author} {\bibfnamefont {A.~A.}\ \bibnamefont {Dzhioev}}\ and\ \bibinfo {author} {\bibfnamefont {D.~S.}\ \bibnamefont {Kosov}},\ }\href {http://aip.scitation.org/doi/full/10.1063/1.3548065} {\bibfield  {journal} {\bibinfo  {journal} {J. Chem. Phys.}\ }\textbf {\bibinfo {volume} {134}},\ \bibinfo {pages} {044121} (\bibinfo {year} {2011})}\BibitemShut {NoStop}%
\bibitem [{\citenamefont {Ajisaka}\ \emph {et~al.}(2012)\citenamefont {Ajisaka}, \citenamefont {Barra}, \citenamefont {Mejía-Monasterio},\ and\ \citenamefont {Prosen}}]{ajisaka_nonequilibrium_2012}%
  \BibitemOpen
  \bibfield  {author} {\bibinfo {author} {\bibfnamefont {S.}~\bibnamefont {Ajisaka}}, \bibinfo {author} {\bibfnamefont {F.}~\bibnamefont {Barra}}, \bibinfo {author} {\bibfnamefont {C.}~\bibnamefont {Mejía-Monasterio}}, \ and\ \bibinfo {author} {\bibfnamefont {T.}~\bibnamefont {Prosen}},\ }\href {http://link.aps.org/doi/10.1103/PhysRevB.86.125111} {\bibfield  {journal} {\bibinfo  {journal} {Phys. Rev. B}\ }\textbf {\bibinfo {volume} {86}},\ \bibinfo {pages} {125111} (\bibinfo {year} {2012})}\BibitemShut {NoStop}%
\bibitem [{\citenamefont {Ajisaka}\ and\ \citenamefont {Barra}(2013)}]{ajisaka_nonequilibrium_2013}%
  \BibitemOpen
  \bibfield  {author} {\bibinfo {author} {\bibfnamefont {S.}~\bibnamefont {Ajisaka}}\ and\ \bibinfo {author} {\bibfnamefont {F.}~\bibnamefont {Barra}},\ }\href {https://link.aps.org/doi/10.1103/PhysRevB.87.195114} {\bibfield  {journal} {\bibinfo  {journal} {Phys. Rev. B}\ }\textbf {\bibinfo {volume} {87}},\ \bibinfo {pages} {195114} (\bibinfo {year} {2013})}\BibitemShut {NoStop}%
\bibitem [{\citenamefont {Chen}\ \emph {et~al.}(2014)\citenamefont {Chen}, \citenamefont {Hansen},\ and\ \citenamefont {Franco}}]{chen_simple_2014}%
  \BibitemOpen
  \bibfield  {author} {\bibinfo {author} {\bibfnamefont {L.}~\bibnamefont {Chen}}, \bibinfo {author} {\bibfnamefont {T.}~\bibnamefont {Hansen}}, \ and\ \bibinfo {author} {\bibfnamefont {I.}~\bibnamefont {Franco}},\ }\href {https://pubs.acs.org/doi/10.1021/jp505771f} {\bibfield  {journal} {\bibinfo  {journal} {J. Phys. Chem. C}\ }\textbf {\bibinfo {volume} {118}},\ \bibinfo {pages} {20009} (\bibinfo {year} {2014})}\BibitemShut {NoStop}%
\bibitem [{\citenamefont {Schwarz}\ \emph {et~al.}(2016)\citenamefont {Schwarz}, \citenamefont {Goldstein}, \citenamefont {Dorda}, \citenamefont {Arrigoni}, \citenamefont {Weichselbaum},\ and\ \citenamefont {von Delft}}]{schwarz_lindblad-driven_2016}%
  \BibitemOpen
  \bibfield  {author} {\bibinfo {author} {\bibfnamefont {F.}~\bibnamefont {Schwarz}}, \bibinfo {author} {\bibfnamefont {M.}~\bibnamefont {Goldstein}}, \bibinfo {author} {\bibfnamefont {A.}~\bibnamefont {Dorda}}, \bibinfo {author} {\bibfnamefont {E.}~\bibnamefont {Arrigoni}}, \bibinfo {author} {\bibfnamefont {A.}~\bibnamefont {Weichselbaum}}, \ and\ \bibinfo {author} {\bibfnamefont {J.}~\bibnamefont {von Delft}},\ }\href {https://link.aps.org/doi/10.1103/PhysRevB.94.155142} {\bibfield  {journal} {\bibinfo  {journal} {Phys. Rev. B}\ }\textbf {\bibinfo {volume} {94}},\ \bibinfo {pages} {155142} (\bibinfo {year} {2016})}\BibitemShut {NoStop}%
\bibitem [{\citenamefont {Chiang}\ and\ \citenamefont {Hsu}(2020)}]{chiang_quantum_2020}%
  \BibitemOpen
  \bibfield  {author} {\bibinfo {author} {\bibfnamefont {T.-M.}\ \bibnamefont {Chiang}}\ and\ \bibinfo {author} {\bibfnamefont {L.-Y.}\ \bibnamefont {Hsu}},\ }\href {https://aip.scitation.org/doi/10.1063/5.0007750} {\bibfield  {journal} {\bibinfo  {journal} {J. Chem. Phys.}\ }\textbf {\bibinfo {volume} {153}},\ \bibinfo {pages} {044103} (\bibinfo {year} {2020})}\BibitemShut {NoStop}%
\bibitem [{\citenamefont {Chen}\ \emph {et~al.}(2019{\natexlab{a}})\citenamefont {Chen}, \citenamefont {Cohen},\ and\ \citenamefont {Galperin}}]{chen_auxiliary_2019}%
  \BibitemOpen
  \bibfield  {author} {\bibinfo {author} {\bibfnamefont {F.}~\bibnamefont {Chen}}, \bibinfo {author} {\bibfnamefont {G.}~\bibnamefont {Cohen}}, \ and\ \bibinfo {author} {\bibfnamefont {M.}~\bibnamefont {Galperin}},\ }\href {https://link.aps.org/doi/10.1103/PhysRevLett.122.186803} {\bibfield  {journal} {\bibinfo  {journal} {Phys. Rev. Lett.}\ }\textbf {\bibinfo {volume} {122}},\ \bibinfo {pages} {186803} (\bibinfo {year} {2019}{\natexlab{a}})}\BibitemShut {NoStop}%
\bibitem [{\citenamefont {Chen}\ \emph {et~al.}(2019{\natexlab{b}})\citenamefont {Chen}, \citenamefont {Arrigoni},\ and\ \citenamefont {Galperin}}]{chen_markovian_2019}%
  \BibitemOpen
  \bibfield  {author} {\bibinfo {author} {\bibfnamefont {F.}~\bibnamefont {Chen}}, \bibinfo {author} {\bibfnamefont {E.}~\bibnamefont {Arrigoni}}, \ and\ \bibinfo {author} {\bibfnamefont {M.}~\bibnamefont {Galperin}},\ }\href {https://doi.org/10.1088/1367-2630/ab5ec5} {\bibfield  {journal} {\bibinfo  {journal} {New J. Phys.}\ }\textbf {\bibinfo {volume} {21}},\ \bibinfo {pages} {123035} (\bibinfo {year} {2019}{\natexlab{b}})}\BibitemShut {NoStop}%
\bibitem [{\citenamefont {Velizhanin}\ \emph {et~al.}(2011)\citenamefont {Velizhanin}, \citenamefont {Chien}, \citenamefont {Dubi},\ and\ \citenamefont {Zwolak}}]{velizhanin_driving_2011}%
  \BibitemOpen
  \bibfield  {author} {\bibinfo {author} {\bibfnamefont {K.~A.}\ \bibnamefont {Velizhanin}}, \bibinfo {author} {\bibfnamefont {C.-C.}\ \bibnamefont {Chien}}, \bibinfo {author} {\bibfnamefont {Y.}~\bibnamefont {Dubi}}, \ and\ \bibinfo {author} {\bibfnamefont {M.}~\bibnamefont {Zwolak}},\ }\href {https://link.aps.org/doi/10.1103/PhysRevE.83.050906} {\bibfield  {journal} {\bibinfo  {journal} {Phys. Rev. E}\ }\textbf {\bibinfo {volume} {83}},\ \bibinfo {pages} {050906} (\bibinfo {year} {2011})}\BibitemShut {NoStop}%
\bibitem [{\citenamefont {Chien}\ \emph {et~al.}(2013{\natexlab{b}})\citenamefont {Chien}, \citenamefont {Velizhanin}, \citenamefont {Dubi},\ and\ \citenamefont {Zwolak}}]{chien_tunable_2013}%
  \BibitemOpen
  \bibfield  {author} {\bibinfo {author} {\bibfnamefont {C.-C.}\ \bibnamefont {Chien}}, \bibinfo {author} {\bibfnamefont {K.~A.}\ \bibnamefont {Velizhanin}}, \bibinfo {author} {\bibfnamefont {Y.}~\bibnamefont {Dubi}}, \ and\ \bibinfo {author} {\bibfnamefont {M.}~\bibnamefont {Zwolak}},\ }\href {https://iopscience.iop.org/article/10.1088/0957-4484/24/9/095704} {\bibfield  {journal} {\bibinfo  {journal} {Nanotechnology}\ }\textbf {\bibinfo {volume} {24}},\ \bibinfo {pages} {095704} (\bibinfo {year} {2013}{\natexlab{b}})}\BibitemShut {NoStop}%
\bibitem [{\citenamefont {Chien}\ \emph {et~al.}(2017)\citenamefont {Chien}, \citenamefont {Kouachi}, \citenamefont {Velizhanin}, \citenamefont {Dubi},\ and\ \citenamefont {Zwolak}}]{chien_thermal_2017}%
  \BibitemOpen
  \bibfield  {author} {\bibinfo {author} {\bibfnamefont {C.-C.}\ \bibnamefont {Chien}}, \bibinfo {author} {\bibfnamefont {S.}~\bibnamefont {Kouachi}}, \bibinfo {author} {\bibfnamefont {K.~A.}\ \bibnamefont {Velizhanin}}, \bibinfo {author} {\bibfnamefont {Y.}~\bibnamefont {Dubi}}, \ and\ \bibinfo {author} {\bibfnamefont {M.}~\bibnamefont {Zwolak}},\ }\href {http://link.aps.org/doi/10.1103/PhysRevE.95.012137} {\bibfield  {journal} {\bibinfo  {journal} {Phys. Rev. E}\ }\textbf {\bibinfo {volume} {95}},\ \bibinfo {pages} {012137} (\bibinfo {year} {2017})}\BibitemShut {NoStop}%
\bibitem [{\citenamefont {Chien}\ \emph {et~al.}(2018)\citenamefont {Chien}, \citenamefont {Velizhanin}, \citenamefont {Dubi}, \citenamefont {Ilic},\ and\ \citenamefont {Zwolak}}]{chien_topological_2018}%
  \BibitemOpen
  \bibfield  {author} {\bibinfo {author} {\bibfnamefont {C.-C.}\ \bibnamefont {Chien}}, \bibinfo {author} {\bibfnamefont {K.~A.}\ \bibnamefont {Velizhanin}}, \bibinfo {author} {\bibfnamefont {Y.}~\bibnamefont {Dubi}}, \bibinfo {author} {\bibfnamefont {B.~R.}\ \bibnamefont {Ilic}}, \ and\ \bibinfo {author} {\bibfnamefont {M.}~\bibnamefont {Zwolak}},\ }\href {https://link.aps.org/doi/10.1103/PhysRevB.97.125425} {\bibfield  {journal} {\bibinfo  {journal} {Phys. Rev. B}\ }\textbf {\bibinfo {volume} {97}},\ \bibinfo {pages} {125425} (\bibinfo {year} {2018})}\BibitemShut {NoStop}%
\bibitem [{\citenamefont {Lacerda}\ \emph {et~al.}(2024)\citenamefont {Lacerda}, \citenamefont {Kewming}, \citenamefont {Brenes}, \citenamefont {Jackson}, \citenamefont {Clark}, \citenamefont {Mitchison},\ and\ \citenamefont {Goold}}]{lacerda_entropy_2024}%
  \BibitemOpen
  \bibfield  {author} {\bibinfo {author} {\bibfnamefont {A.~M.}\ \bibnamefont {Lacerda}}, \bibinfo {author} {\bibfnamefont {M.~J.}\ \bibnamefont {Kewming}}, \bibinfo {author} {\bibfnamefont {M.}~\bibnamefont {Brenes}}, \bibinfo {author} {\bibfnamefont {C.}~\bibnamefont {Jackson}}, \bibinfo {author} {\bibfnamefont {S.~R.}\ \bibnamefont {Clark}}, \bibinfo {author} {\bibfnamefont {M.~T.}\ \bibnamefont {Mitchison}}, \ and\ \bibinfo {author} {\bibfnamefont {J.}~\bibnamefont {Goold}},\ }\href {https://link.aps.org/doi/10.1103/PhysRevE.110.014125} {\bibfield  {journal} {\bibinfo  {journal} {Phys. Rev. E}\ }\textbf {\bibinfo {volume} {110}},\ \bibinfo {pages} {014125} (\bibinfo {year} {2024})}\BibitemShut {NoStop}%
\bibitem [{\citenamefont {Nazir}\ and\ \citenamefont {Schaller}(2018)}]{binder_reaction_2018}%
  \BibitemOpen
  \bibfield  {author} {\bibinfo {author} {\bibfnamefont {A.}~\bibnamefont {Nazir}}\ and\ \bibinfo {author} {\bibfnamefont {G.}~\bibnamefont {Schaller}},\ }in\ \href {http://link.springer.com/10.1007/978-3-319-99046-0_23} {\emph {\bibinfo {booktitle} {Thermodynamics in the {Quantum} {Regime}}}},\ Vol.\ \bibinfo {volume} {195},\ \bibinfo {editor} {edited by\ \bibinfo {editor} {\bibfnamefont {F.}~\bibnamefont {Binder}}, \bibinfo {editor} {\bibfnamefont {L.~A.}\ \bibnamefont {Correa}}, \bibinfo {editor} {\bibfnamefont {C.}~\bibnamefont {Gogolin}}, \bibinfo {editor} {\bibfnamefont {J.}~\bibnamefont {Anders}}, \ and\ \bibinfo {editor} {\bibfnamefont {G.}~\bibnamefont {Adesso}}}\ (\bibinfo  {publisher} {Springer International Publishing},\ \bibinfo {address} {Cham},\ \bibinfo {year} {2018})\ pp.\ \bibinfo {pages} {551--577}\BibitemShut {NoStop}%
\bibitem [{\citenamefont {Dast}\ \emph {et~al.}(2014)\citenamefont {Dast}, \citenamefont {Haag}, \citenamefont {Cartarius},\ and\ \citenamefont {Wunner}}]{dast_quantum_2014}%
  \BibitemOpen
  \bibfield  {author} {\bibinfo {author} {\bibfnamefont {D.}~\bibnamefont {Dast}}, \bibinfo {author} {\bibfnamefont {D.}~\bibnamefont {Haag}}, \bibinfo {author} {\bibfnamefont {H.}~\bibnamefont {Cartarius}}, \ and\ \bibinfo {author} {\bibfnamefont {G.}~\bibnamefont {Wunner}},\ }\href {https://link.aps.org/doi/10.1103/PhysRevA.90.052120} {\bibfield  {journal} {\bibinfo  {journal} {Phys. Rev. A}\ }\textbf {\bibinfo {volume} {90}},\ \bibinfo {pages} {052120} (\bibinfo {year} {2014})}\BibitemShut {NoStop}%
\bibitem [{\citenamefont {Ajisaka}\ \emph {et~al.}(2015)\citenamefont {Ajisaka}, \citenamefont {Žunkovič},\ and\ \citenamefont {Dubi}}]{ajisaka_molecular_2015}%
  \BibitemOpen
  \bibfield  {author} {\bibinfo {author} {\bibfnamefont {S.}~\bibnamefont {Ajisaka}}, \bibinfo {author} {\bibfnamefont {B.}~\bibnamefont {Žunkovič}}, \ and\ \bibinfo {author} {\bibfnamefont {Y.}~\bibnamefont {Dubi}},\ }\href {https://www.nature.com/articles/srep08312} {\bibfield  {journal} {\bibinfo  {journal} {Sci. Rep.}\ }\textbf {\bibinfo {volume} {5}},\ \bibinfo {pages} {8312} (\bibinfo {year} {2015})}\BibitemShut {NoStop}%
\bibitem [{\citenamefont {Mahajan}\ \emph {et~al.}(2016)\citenamefont {Mahajan}, \citenamefont {Freeman}, \citenamefont {Mumford}, \citenamefont {Tubman},\ and\ \citenamefont {Swingle}}]{mahajan_entanglement_2016}%
  \BibitemOpen
  \bibfield  {author} {\bibinfo {author} {\bibfnamefont {R.}~\bibnamefont {Mahajan}}, \bibinfo {author} {\bibfnamefont {C.~D.}\ \bibnamefont {Freeman}}, \bibinfo {author} {\bibfnamefont {S.}~\bibnamefont {Mumford}}, \bibinfo {author} {\bibfnamefont {N.}~\bibnamefont {Tubman}}, \ and\ \bibinfo {author} {\bibfnamefont {B.}~\bibnamefont {Swingle}},\ }\href {http://arxiv.org/abs/1608.05074} {\bibfield  {journal} {\bibinfo  {journal} {arXiv:1608.05074}\ } (\bibinfo {year} {2016})}\BibitemShut {NoStop}%
\bibitem [{\citenamefont {Zanoci}\ and\ \citenamefont {Swingle}(2020)}]{zanoci_entanglement_2020}%
  \BibitemOpen
  \bibfield  {author} {\bibinfo {author} {\bibfnamefont {C.}~\bibnamefont {Zanoci}}\ and\ \bibinfo {author} {\bibfnamefont {B.}~\bibnamefont {Swingle}},\ }\href {http://arxiv.org/abs/1612.04840} {\bibfield  {journal} {\bibinfo  {journal} {arXiv:1612.04840}\ } (\bibinfo {year} {2020})}\BibitemShut {NoStop}%
\bibitem [{\citenamefont {Tamascelli}\ \emph {et~al.}(2018)\citenamefont {Tamascelli}, \citenamefont {Smirne}, \citenamefont {Huelga},\ and\ \citenamefont {Plenio}}]{tamascelli_nonperturbative_2018}%
  \BibitemOpen
  \bibfield  {author} {\bibinfo {author} {\bibfnamefont {D.}~\bibnamefont {Tamascelli}}, \bibinfo {author} {\bibfnamefont {A.}~\bibnamefont {Smirne}}, \bibinfo {author} {\bibfnamefont {S.}~\bibnamefont {Huelga}}, \ and\ \bibinfo {author} {\bibfnamefont {M.}~\bibnamefont {Plenio}},\ }\href {https://link.aps.org/doi/10.1103/PhysRevLett.120.030402} {\bibfield  {journal} {\bibinfo  {journal} {Phys. Rev. Lett.}\ }\textbf {\bibinfo {volume} {120}},\ \bibinfo {pages} {030402} (\bibinfo {year} {2018})}\BibitemShut {NoStop}%
\bibitem [{\citenamefont {Imamog¯lu}(1994)}]{imamoglu_stochastic_1994}%
  \BibitemOpen
  \bibfield  {author} {\bibinfo {author} {\bibfnamefont {A.}~\bibnamefont {Imamog¯lu}},\ }\href {https://link.aps.org/doi/10.1103/PhysRevA.50.3650} {\bibfield  {journal} {\bibinfo  {journal} {Phys. Rev. A}\ }\textbf {\bibinfo {volume} {50}},\ \bibinfo {pages} {3650} (\bibinfo {year} {1994})}\BibitemShut {NoStop}%
\bibitem [{\citenamefont {Garraway}(1997{\natexlab{a}})}]{garraway_nonperturbative_1997}%
  \BibitemOpen
  \bibfield  {author} {\bibinfo {author} {\bibfnamefont {B.~M.}\ \bibnamefont {Garraway}},\ }\href {https://link.aps.org/doi/10.1103/PhysRevA.55.2290} {\bibfield  {journal} {\bibinfo  {journal} {Phys. Rev. A}\ }\textbf {\bibinfo {volume} {55}},\ \bibinfo {pages} {2290} (\bibinfo {year} {1997}{\natexlab{a}})}\BibitemShut {NoStop}%
\bibitem [{\citenamefont {Garraway}(1997{\natexlab{b}})}]{garraway_decay_1997}%
  \BibitemOpen
  \bibfield  {author} {\bibinfo {author} {\bibfnamefont {B.~M.}\ \bibnamefont {Garraway}},\ }\href {https://link.aps.org/doi/10.1103/PhysRevA.55.4636} {\bibfield  {journal} {\bibinfo  {journal} {Phys. Rev. A}\ }\textbf {\bibinfo {volume} {55}},\ \bibinfo {pages} {4636} (\bibinfo {year} {1997}{\natexlab{b}})}\BibitemShut {NoStop}%
\bibitem [{\citenamefont {Zwolak}(2008{\natexlab{b}})}]{zwolak_dynamics_2008}%
  \BibitemOpen
  \bibfield  {author} {\bibinfo {author} {\bibfnamefont {M.}~\bibnamefont {Zwolak}},\ }\emph {\bibinfo {title} {Dynamics and {Simulation} of {Open} {Quantum} {Systems}}},\ \href {https://resolver.caltech.edu/CaltechETD:etd-08072007-221313} {\bibinfo {type} {{PhD}}},\ \bibinfo  {school} {California Institute of Technology} (\bibinfo {year} {2008}{\natexlab{b}})\BibitemShut {NoStop}%
\bibitem [{\citenamefont {Pleasance}\ \emph {et~al.}(2020)\citenamefont {Pleasance}, \citenamefont {Garraway},\ and\ \citenamefont {Petruccione}}]{pleasance_generalized_2020}%
  \BibitemOpen
  \bibfield  {author} {\bibinfo {author} {\bibfnamefont {G.}~\bibnamefont {Pleasance}}, \bibinfo {author} {\bibfnamefont {B.~M.}\ \bibnamefont {Garraway}}, \ and\ \bibinfo {author} {\bibfnamefont {F.}~\bibnamefont {Petruccione}},\ }\href {https://link.aps.org/doi/10.1103/PhysRevResearch.2.043058} {\bibfield  {journal} {\bibinfo  {journal} {Phys. Rev. Res.}\ }\textbf {\bibinfo {volume} {2}},\ \bibinfo {pages} {043058} (\bibinfo {year} {2020})}\BibitemShut {NoStop}%
\bibitem [{\citenamefont {Morzan}\ \emph {et~al.}(2017)\citenamefont {Morzan}, \citenamefont {Ramírez}, \citenamefont {González~Lebrero},\ and\ \citenamefont {Scherlis}}]{morzan_electron_2017}%
  \BibitemOpen
  \bibfield  {author} {\bibinfo {author} {\bibfnamefont {U.~N.}\ \bibnamefont {Morzan}}, \bibinfo {author} {\bibfnamefont {F.~F.}\ \bibnamefont {Ramírez}}, \bibinfo {author} {\bibfnamefont {M.~C.}\ \bibnamefont {González~Lebrero}}, \ and\ \bibinfo {author} {\bibfnamefont {D.~A.}\ \bibnamefont {Scherlis}},\ }\href {https://doi.org/10.1063/1.4974095} {\bibfield  {journal} {\bibinfo  {journal} {J. Chem. Phys.}\ }\textbf {\bibinfo {volume} {146}},\ \bibinfo {pages} {044110} (\bibinfo {year} {2017})}\BibitemShut {NoStop}%
\bibitem [{\citenamefont {Oz}\ \emph {et~al.}(2023)\citenamefont {Oz}, \citenamefont {Nitzan}, \citenamefont {Hod},\ and\ \citenamefont {Peralta}}]{oz_electron_2023}%
  \BibitemOpen
  \bibfield  {author} {\bibinfo {author} {\bibfnamefont {A.}~\bibnamefont {Oz}}, \bibinfo {author} {\bibfnamefont {A.}~\bibnamefont {Nitzan}}, \bibinfo {author} {\bibfnamefont {O.}~\bibnamefont {Hod}}, \ and\ \bibinfo {author} {\bibfnamefont {J.~E.}\ \bibnamefont {Peralta}},\ }\href {https://doi.org/10.1021/acs.jctc.3c00311} {\bibfield  {journal} {\bibinfo  {journal} {J. Chem. Theory Comput.}\ }\textbf {\bibinfo {volume} {19}},\ \bibinfo {pages} {7496} (\bibinfo {year} {2023})}\BibitemShut {NoStop}%
\bibitem [{\citenamefont {Arrigoni}\ \emph {et~al.}(2013)\citenamefont {Arrigoni}, \citenamefont {Knap},\ and\ \citenamefont {von~der Linden}}]{arrigoni_nonequilibrium_2013}%
  \BibitemOpen
  \bibfield  {author} {\bibinfo {author} {\bibfnamefont {E.}~\bibnamefont {Arrigoni}}, \bibinfo {author} {\bibfnamefont {M.}~\bibnamefont {Knap}}, \ and\ \bibinfo {author} {\bibfnamefont {W.}~\bibnamefont {von~der Linden}},\ }\href {https://link.aps.org/doi/10.1103/PhysRevLett.110.086403} {\bibfield  {journal} {\bibinfo  {journal} {Phys. Rev. Lett.}\ }\textbf {\bibinfo {volume} {110}},\ \bibinfo {pages} {086403} (\bibinfo {year} {2013})}\BibitemShut {NoStop}%
\bibitem [{\citenamefont {Dorda}\ \emph {et~al.}(2014)\citenamefont {Dorda}, \citenamefont {Nuss}, \citenamefont {von~der Linden},\ and\ \citenamefont {Arrigoni}}]{dorda_auxiliary_2014}%
  \BibitemOpen
  \bibfield  {author} {\bibinfo {author} {\bibfnamefont {A.}~\bibnamefont {Dorda}}, \bibinfo {author} {\bibfnamefont {M.}~\bibnamefont {Nuss}}, \bibinfo {author} {\bibfnamefont {W.}~\bibnamefont {von~der Linden}}, \ and\ \bibinfo {author} {\bibfnamefont {E.}~\bibnamefont {Arrigoni}},\ }\href {https://link.aps.org/doi/10.1103/PhysRevB.89.165105} {\bibfield  {journal} {\bibinfo  {journal} {Phys. Rev. B}\ }\textbf {\bibinfo {volume} {89}},\ \bibinfo {pages} {165105} (\bibinfo {year} {2014})}\BibitemShut {NoStop}%
\bibitem [{\citenamefont {Dorda}\ \emph {et~al.}(2017)\citenamefont {Dorda}, \citenamefont {Sorantin}, \citenamefont {Linden},\ and\ \citenamefont {Arrigoni}}]{dorda_optimized_2017}%
  \BibitemOpen
  \bibfield  {author} {\bibinfo {author} {\bibfnamefont {A.}~\bibnamefont {Dorda}}, \bibinfo {author} {\bibfnamefont {M.}~\bibnamefont {Sorantin}}, \bibinfo {author} {\bibfnamefont {W.~v.~d.}\ \bibnamefont {Linden}}, \ and\ \bibinfo {author} {\bibfnamefont {E.}~\bibnamefont {Arrigoni}},\ }\href {https://doi.org/10.1088/1367-2630/aa6ccc} {\bibfield  {journal} {\bibinfo  {journal} {New J. Phys.}\ }\textbf {\bibinfo {volume} {19}},\ \bibinfo {pages} {063005} (\bibinfo {year} {2017})}\BibitemShut {NoStop}%
\bibitem [{\citenamefont {Dorda}\ \emph {et~al.}(2015)\citenamefont {Dorda}, \citenamefont {Ganahl}, \citenamefont {Evertz}, \citenamefont {von~der Linden},\ and\ \citenamefont {Arrigoni}}]{dorda_auxiliary_2015}%
  \BibitemOpen
  \bibfield  {author} {\bibinfo {author} {\bibfnamefont {A.}~\bibnamefont {Dorda}}, \bibinfo {author} {\bibfnamefont {M.}~\bibnamefont {Ganahl}}, \bibinfo {author} {\bibfnamefont {H.~G.}\ \bibnamefont {Evertz}}, \bibinfo {author} {\bibfnamefont {W.}~\bibnamefont {von~der Linden}}, \ and\ \bibinfo {author} {\bibfnamefont {E.}~\bibnamefont {Arrigoni}},\ }\href {https://link.aps.org/doi/10.1103/PhysRevB.92.125145} {\bibfield  {journal} {\bibinfo  {journal} {Phys. Rev. B}\ }\textbf {\bibinfo {volume} {92}},\ \bibinfo {pages} {125145} (\bibinfo {year} {2015})}\BibitemShut {NoStop}%
\bibitem [{\citenamefont {Fugger}\ \emph {et~al.}(2018)\citenamefont {Fugger}, \citenamefont {Dorda}, \citenamefont {Schwarz}, \citenamefont {von Delft},\ and\ \citenamefont {Arrigoni}}]{fugger_nonequilibrium_2018}%
  \BibitemOpen
  \bibfield  {author} {\bibinfo {author} {\bibfnamefont {D.~M.}\ \bibnamefont {Fugger}}, \bibinfo {author} {\bibfnamefont {A.}~\bibnamefont {Dorda}}, \bibinfo {author} {\bibfnamefont {F.}~\bibnamefont {Schwarz}}, \bibinfo {author} {\bibfnamefont {J.}~\bibnamefont {von Delft}}, \ and\ \bibinfo {author} {\bibfnamefont {E.}~\bibnamefont {Arrigoni}},\ }\href {https://doi.org/10.1088%2F1367-2630%2Faa9fdc} {\bibfield  {journal} {\bibinfo  {journal} {New J. Phys.}\ }\textbf {\bibinfo {volume} {20}},\ \bibinfo {pages} {013030} (\bibinfo {year} {2018})}\BibitemShut {NoStop}%
\bibitem [{\citenamefont {Fugger}\ \emph {et~al.}(2020)\citenamefont {Fugger}, \citenamefont {Bauernfeind}, \citenamefont {Sorantin},\ and\ \citenamefont {Arrigoni}}]{fugger_nonequilibrium_2020}%
  \BibitemOpen
  \bibfield  {author} {\bibinfo {author} {\bibfnamefont {D.~M.}\ \bibnamefont {Fugger}}, \bibinfo {author} {\bibfnamefont {D.}~\bibnamefont {Bauernfeind}}, \bibinfo {author} {\bibfnamefont {M.~E.}\ \bibnamefont {Sorantin}}, \ and\ \bibinfo {author} {\bibfnamefont {E.}~\bibnamefont {Arrigoni}},\ }\href {https://link.aps.org/doi/10.1103/PhysRevB.101.165132} {\bibfield  {journal} {\bibinfo  {journal} {Phys. Rev. B}\ }\textbf {\bibinfo {volume} {101}},\ \bibinfo {pages} {165132} (\bibinfo {year} {2020})}\BibitemShut {NoStop}%
\bibitem [{\citenamefont {Lotem}\ \emph {et~al.}(2020)\citenamefont {Lotem}, \citenamefont {Weichselbaum}, \citenamefont {von Delft},\ and\ \citenamefont {Goldstein}}]{lotem_renormalized_2020}%
  \BibitemOpen
  \bibfield  {author} {\bibinfo {author} {\bibfnamefont {M.}~\bibnamefont {Lotem}}, \bibinfo {author} {\bibfnamefont {A.}~\bibnamefont {Weichselbaum}}, \bibinfo {author} {\bibfnamefont {J.}~\bibnamefont {von Delft}}, \ and\ \bibinfo {author} {\bibfnamefont {M.}~\bibnamefont {Goldstein}},\ }\href {https://link.aps.org/doi/10.1103/PhysRevResearch.2.043052} {\bibfield  {journal} {\bibinfo  {journal} {Phys. Rev. Res.}\ }\textbf {\bibinfo {volume} {2}},\ \bibinfo {pages} {043052} (\bibinfo {year} {2020})}\BibitemShut {NoStop}%
\bibitem [{\citenamefont {Brenes}\ \emph {et~al.}(2020)\citenamefont {Brenes}, \citenamefont {Mendoza-Arenas}, \citenamefont {Purkayastha}, \citenamefont {Mitchison}, \citenamefont {Clark},\ and\ \citenamefont {Goold}}]{brenes_tensor-network_2020}%
  \BibitemOpen
  \bibfield  {author} {\bibinfo {author} {\bibfnamefont {M.}~\bibnamefont {Brenes}}, \bibinfo {author} {\bibfnamefont {J.~J.}\ \bibnamefont {Mendoza-Arenas}}, \bibinfo {author} {\bibfnamefont {A.}~\bibnamefont {Purkayastha}}, \bibinfo {author} {\bibfnamefont {M.~T.}\ \bibnamefont {Mitchison}}, \bibinfo {author} {\bibfnamefont {S.~R.}\ \bibnamefont {Clark}}, \ and\ \bibinfo {author} {\bibfnamefont {J.}~\bibnamefont {Goold}},\ }\href {https://link.aps.org/doi/10.1103/PhysRevX.10.031040} {\bibfield  {journal} {\bibinfo  {journal} {Phys. Rev. X}\ }\textbf {\bibinfo {volume} {10}},\ \bibinfo {pages} {031040} (\bibinfo {year} {2020})}\BibitemShut {NoStop}%
\bibitem [{\citenamefont {Rams}\ and\ \citenamefont {Zwolak}(2020)}]{rams_breaking_2020}%
  \BibitemOpen
  \bibfield  {author} {\bibinfo {author} {\bibfnamefont {M.~M.}\ \bibnamefont {Rams}}\ and\ \bibinfo {author} {\bibfnamefont {M.}~\bibnamefont {Zwolak}},\ }\href {https://link.aps.org/doi/10.1103/PhysRevLett.124.137701} {\bibfield  {journal} {\bibinfo  {journal} {Phys. Rev. Lett.}\ }\textbf {\bibinfo {volume} {124}},\ \bibinfo {pages} {137701} (\bibinfo {year} {2020})}\BibitemShut {NoStop}%
\bibitem [{\citenamefont {W\'{o}jtowicz}\ \emph {et~al.}(2020)\citenamefont {W\'{o}jtowicz}, \citenamefont {Elenewski}, \citenamefont {Rams},\ and\ \citenamefont {Zwolak}}]{wojtowicz_open-system_2020}%
  \BibitemOpen
  \bibfield  {author} {\bibinfo {author} {\bibfnamefont {G.}~\bibnamefont {W\'{o}jtowicz}}, \bibinfo {author} {\bibfnamefont {J.~E.}\ \bibnamefont {Elenewski}}, \bibinfo {author} {\bibfnamefont {M.~M.}\ \bibnamefont {Rams}}, \ and\ \bibinfo {author} {\bibfnamefont {M.}~\bibnamefont {Zwolak}},\ }\href {https://link.aps.org/doi/10.1103/PhysRevA.101.050301} {\bibfield  {journal} {\bibinfo  {journal} {Phys. Rev. A}\ }\textbf {\bibinfo {volume} {101}},\ \bibinfo {pages} {050301} (\bibinfo {year} {2020})}\BibitemShut {NoStop}%
\bibitem [{\citenamefont {Verstraete}\ \emph {et~al.}(2004)\citenamefont {Verstraete}, \citenamefont {García-Ripoll},\ and\ \citenamefont {Cirac}}]{verstraete_matrix_2004}%
  \BibitemOpen
  \bibfield  {author} {\bibinfo {author} {\bibfnamefont {F.}~\bibnamefont {Verstraete}}, \bibinfo {author} {\bibfnamefont {J.~J.}\ \bibnamefont {García-Ripoll}}, \ and\ \bibinfo {author} {\bibfnamefont {J.~I.}\ \bibnamefont {Cirac}},\ }\href {https://link.aps.org/doi/10.1103/PhysRevLett.93.207204} {\bibfield  {journal} {\bibinfo  {journal} {Phys. Rev. Lett.}\ }\textbf {\bibinfo {volume} {93}},\ \bibinfo {pages} {207204} (\bibinfo {year} {2004})}\BibitemShut {NoStop}%
\bibitem [{\citenamefont {De}\ \emph {et~al.}(2023)\citenamefont {De}, \citenamefont {Wójtowicz}, \citenamefont {Zakrzewski}, \citenamefont {Zwolak},\ and\ \citenamefont {Rams}}]{de_transport_2023}%
  \BibitemOpen
  \bibfield  {author} {\bibinfo {author} {\bibfnamefont {B.}~\bibnamefont {De}}, \bibinfo {author} {\bibfnamefont {G.}~\bibnamefont {Wójtowicz}}, \bibinfo {author} {\bibfnamefont {J.}~\bibnamefont {Zakrzewski}}, \bibinfo {author} {\bibfnamefont {M.}~\bibnamefont {Zwolak}}, \ and\ \bibinfo {author} {\bibfnamefont {M.~M.}\ \bibnamefont {Rams}},\ }\href {https://link.aps.org/doi/10.1103/PhysRevB.107.235148} {\bibfield  {journal} {\bibinfo  {journal} {Phys. Rev. B}\ }\textbf {\bibinfo {volume} {107}},\ \bibinfo {pages} {235148} (\bibinfo {year} {2023})}\BibitemShut {NoStop}%
\bibitem [{\citenamefont {De}\ \emph {et~al.}(2024)\citenamefont {De}, \citenamefont {Wójtowicz}, \citenamefont {Rams}, \citenamefont {Zwolak},\ and\ \citenamefont {Zakrzewski}}]{de_confluence_2024}%
  \BibitemOpen
  \bibfield  {author} {\bibinfo {author} {\bibfnamefont {B.}~\bibnamefont {De}}, \bibinfo {author} {\bibfnamefont {G.}~\bibnamefont {Wójtowicz}}, \bibinfo {author} {\bibfnamefont {M.~M.}\ \bibnamefont {Rams}}, \bibinfo {author} {\bibfnamefont {M.}~\bibnamefont {Zwolak}}, \ and\ \bibinfo {author} {\bibfnamefont {J.}~\bibnamefont {Zakrzewski}},\ }\href {https://link.aps.org/doi/10.1103/PhysRevB.110.155146} {\bibfield  {journal} {\bibinfo  {journal} {Phys. Rev. B}\ }\textbf {\bibinfo {volume} {110}},\ \bibinfo {pages} {155146} (\bibinfo {year} {2024})}\BibitemShut {NoStop}%
\bibitem [{\citenamefont {Sarkar}\ and\ \citenamefont {Dubi}(2022{\natexlab{a}})}]{sarkar_emergence_2022}%
  \BibitemOpen
  \bibfield  {author} {\bibinfo {author} {\bibfnamefont {S.}~\bibnamefont {Sarkar}}\ and\ \bibinfo {author} {\bibfnamefont {Y.}~\bibnamefont {Dubi}},\ }\href {https://pubs.acs.org/doi/10.1021/acs.nanolett.2c00976} {\bibfield  {journal} {\bibinfo  {journal} {Nano Lett.}\ }\textbf {\bibinfo {volume} {22}},\ \bibinfo {pages} {4445} (\bibinfo {year} {2022}{\natexlab{a}})}\BibitemShut {NoStop}%
\bibitem [{\citenamefont {Sarkar}\ and\ \citenamefont {Dubi}(2022{\natexlab{b}})}]{sarkar_signatures_2022}%
  \BibitemOpen
  \bibfield  {author} {\bibinfo {author} {\bibfnamefont {S.}~\bibnamefont {Sarkar}}\ and\ \bibinfo {author} {\bibfnamefont {Y.}~\bibnamefont {Dubi}},\ }\href {https://www.nature.com/articles/s42005-022-00925-z} {\bibfield  {journal} {\bibinfo  {journal} {Commun. Phys.}\ }\textbf {\bibinfo {volume} {5}},\ \bibinfo {pages} {1} (\bibinfo {year} {2022}{\natexlab{b}})}\BibitemShut {NoStop}%
\bibitem [{\citenamefont {Landi}\ \emph {et~al.}(2022)\citenamefont {Landi}, \citenamefont {Poletti},\ and\ \citenamefont {Schaller}}]{landi_nonequilibrium_2022}%
  \BibitemOpen
  \bibfield  {author} {\bibinfo {author} {\bibfnamefont {G.~T.}\ \bibnamefont {Landi}}, \bibinfo {author} {\bibfnamefont {D.}~\bibnamefont {Poletti}}, \ and\ \bibinfo {author} {\bibfnamefont {G.}~\bibnamefont {Schaller}},\ }\href {https://link.aps.org/doi/10.1103/RevModPhys.94.045006} {\bibfield  {journal} {\bibinfo  {journal} {Rev. Mod. Phys.}\ }\textbf {\bibinfo {volume} {94}},\ \bibinfo {pages} {045006} (\bibinfo {year} {2022})}\BibitemShut {NoStop}%
\bibitem [{\citenamefont {Dhawan}\ \emph {et~al.}(2024)\citenamefont {Dhawan}, \citenamefont {Ganguly}, \citenamefont {Kulkarni},\ and\ \citenamefont {Agarwalla}}]{dhawan_anomalous_2024}%
  \BibitemOpen
  \bibfield  {author} {\bibinfo {author} {\bibfnamefont {A.}~\bibnamefont {Dhawan}}, \bibinfo {author} {\bibfnamefont {K.}~\bibnamefont {Ganguly}}, \bibinfo {author} {\bibfnamefont {M.}~\bibnamefont {Kulkarni}}, \ and\ \bibinfo {author} {\bibfnamefont {B.~K.}\ \bibnamefont {Agarwalla}},\ }\href {https://link.aps.org/doi/10.1103/PhysRevB.110.L081403} {\bibfield  {journal} {\bibinfo  {journal} {Phys. Rev. B}\ }\textbf {\bibinfo {volume} {110}},\ \bibinfo {pages} {L081403} (\bibinfo {year} {2024})}\BibitemShut {NoStop}%
\bibitem [{\citenamefont {Sarkar}\ \emph {et~al.}(2024)\citenamefont {Sarkar}, \citenamefont {Agarwalla},\ and\ \citenamefont {Bhakuni}}]{sarkar_impact_2024}%
  \BibitemOpen
  \bibfield  {author} {\bibinfo {author} {\bibfnamefont {S.}~\bibnamefont {Sarkar}}, \bibinfo {author} {\bibfnamefont {B.~K.}\ \bibnamefont {Agarwalla}}, \ and\ \bibinfo {author} {\bibfnamefont {D.~S.}\ \bibnamefont {Bhakuni}},\ }\href {https://link.aps.org/doi/10.1103/PhysRevB.109.165408} {\bibfield  {journal} {\bibinfo  {journal} {Phys. Rev. B}\ }\textbf {\bibinfo {volume} {109}},\ \bibinfo {pages} {165408} (\bibinfo {year} {2024})}\BibitemShut {NoStop}%
\bibitem [{\citenamefont {Tater}\ \emph {et~al.}(2025)\citenamefont {Tater}, \citenamefont {Sarkar}, \citenamefont {Bhakuni},\ and\ \citenamefont {Agarwalla}}]{tater_bipartite_2025}%
  \BibitemOpen
  \bibfield  {author} {\bibinfo {author} {\bibfnamefont {L.}~\bibnamefont {Tater}}, \bibinfo {author} {\bibfnamefont {S.}~\bibnamefont {Sarkar}}, \bibinfo {author} {\bibfnamefont {D.~S.}\ \bibnamefont {Bhakuni}}, \ and\ \bibinfo {author} {\bibfnamefont {B.~K.}\ \bibnamefont {Agarwalla}},\ }\href {http://arxiv.org/abs/2503.20356} {\bibfield  {journal} {\bibinfo  {journal} {arXiv:2503.20356}\ } (\bibinfo {year} {2025})}\BibitemShut {NoStop}%
\bibitem [{\citenamefont {Lindblad}(1976)}]{lindblad_generators_1976}%
  \BibitemOpen
  \bibfield  {author} {\bibinfo {author} {\bibfnamefont {G.}~\bibnamefont {Lindblad}},\ }\href {https://doi.org/10.1007/BF01608499} {\bibfield  {journal} {\bibinfo  {journal} {Commun. Math. Phys.}\ }\textbf {\bibinfo {volume} {48}},\ \bibinfo {pages} {119} (\bibinfo {year} {1976})}\BibitemShut {NoStop}%
\bibitem [{\citenamefont {Gorini}\ \emph {et~al.}(1976)\citenamefont {Gorini}, \citenamefont {Kossakowski},\ and\ \citenamefont {Sudarshan}}]{gorini_completely_1976}%
  \BibitemOpen
  \bibfield  {author} {\bibinfo {author} {\bibfnamefont {V.}~\bibnamefont {Gorini}}, \bibinfo {author} {\bibfnamefont {A.}~\bibnamefont {Kossakowski}}, \ and\ \bibinfo {author} {\bibfnamefont {E.~C.~G.}\ \bibnamefont {Sudarshan}},\ }\href {https://doi.org/10.1063/1.522979} {\bibfield  {journal} {\bibinfo  {journal} {J. Math. Phys.}\ }\textbf {\bibinfo {volume} {17}},\ \bibinfo {pages} {821} (\bibinfo {year} {1976})}\BibitemShut {NoStop}%
\bibitem [{\citenamefont {Breuer}\ and\ \citenamefont {Petruccione}(2007)}]{breuer_theory_2007}%
  \BibitemOpen
  \bibfield  {author} {\bibinfo {author} {\bibfnamefont {H.-P.}\ \bibnamefont {Breuer}}\ and\ \bibinfo {author} {\bibfnamefont {F.}~\bibnamefont {Petruccione}},\ }\href {https://academic.oup.com/book/27757} {\emph {\bibinfo {title} {The {Theory} of {Open} {Quantum} {Systems}}}},\ \bibinfo {edition} {1st}\ ed.\ (\bibinfo  {publisher} {Oxford University Press, Oxford},\ \bibinfo {year} {2007})\BibitemShut {NoStop}%
\bibitem [{\citenamefont {Bartels}\ and\ \citenamefont {Stewart}(1972)}]{bartels_algorithm_1972}%
  \BibitemOpen
  \bibfield  {author} {\bibinfo {author} {\bibfnamefont {R.~H.}\ \bibnamefont {Bartels}}\ and\ \bibinfo {author} {\bibfnamefont {G.~W.}\ \bibnamefont {Stewart}},\ }\href {https://dl.acm.org/doi/10.1145/361573.361582} {\bibfield  {journal} {\bibinfo  {journal} {Commun. ACM}\ }\textbf {\bibinfo {volume} {15}},\ \bibinfo {pages} {820} (\bibinfo {year} {1972})}\BibitemShut {NoStop}%
\bibitem [{\citenamefont {Golub}\ \emph {et~al.}(1979)\citenamefont {Golub}, \citenamefont {Nash},\ and\ \citenamefont {Van~Loan}}]{golub_hessenberg-schur_1979}%
  \BibitemOpen
  \bibfield  {author} {\bibinfo {author} {\bibfnamefont {G.}~\bibnamefont {Golub}}, \bibinfo {author} {\bibfnamefont {S.}~\bibnamefont {Nash}}, \ and\ \bibinfo {author} {\bibfnamefont {C.}~\bibnamefont {Van~Loan}},\ }\href {https://ieeexplore.ieee.org/abstract/document/1102170} {\bibfield  {journal} {\bibinfo  {journal} {IEEE Trans. Autom. Control.}\ }\textbf {\bibinfo {volume} {24}},\ \bibinfo {pages} {909} (\bibinfo {year} {1979})}\BibitemShut {NoStop}%
\bibitem [{\citenamefont {Turkeshi}\ and\ \citenamefont {Schiró}(2021)}]{turkeshi_diffusion_2021}%
  \BibitemOpen
  \bibfield  {author} {\bibinfo {author} {\bibfnamefont {X.}~\bibnamefont {Turkeshi}}\ and\ \bibinfo {author} {\bibfnamefont {M.}~\bibnamefont {Schiró}},\ }\href {https://link.aps.org/doi/10.1103/PhysRevB.104.144301} {\bibfield  {journal} {\bibinfo  {journal} {Phys. Rev. B}\ }\textbf {\bibinfo {volume} {104}},\ \bibinfo {pages} {144301} (\bibinfo {year} {2021})}\BibitemShut {NoStop}%
\bibitem [{\citenamefont {Varma}\ \emph {et~al.}(2017)\citenamefont {Varma}, \citenamefont {de~Mulatier},\ and\ \citenamefont {Žnidarič}}]{varma_fractality_2017}%
  \BibitemOpen
  \bibfield  {author} {\bibinfo {author} {\bibfnamefont {V.~K.}\ \bibnamefont {Varma}}, \bibinfo {author} {\bibfnamefont {C.}~\bibnamefont {de~Mulatier}}, \ and\ \bibinfo {author} {\bibfnamefont {M.}~\bibnamefont {Žnidarič}},\ }\href {https://link.aps.org/doi/10.1103/PhysRevE.96.032130} {\bibfield  {journal} {\bibinfo  {journal} {Phys. Rev. E}\ }\textbf {\bibinfo {volume} {96}},\ \bibinfo {pages} {032130} (\bibinfo {year} {2017})}\BibitemShut {NoStop}%
\bibitem [{\citenamefont {Taylor}\ and\ \citenamefont {Scardicchio}(2021)}]{taylor_subdiffusion_2021}%
  \BibitemOpen
  \bibfield  {author} {\bibinfo {author} {\bibfnamefont {S.~R.}\ \bibnamefont {Taylor}}\ and\ \bibinfo {author} {\bibfnamefont {A.}~\bibnamefont {Scardicchio}},\ }\href {https://link.aps.org/doi/10.1103/PhysRevB.103.184202} {\bibfield  {journal} {\bibinfo  {journal} {Phys. Rev. B}\ }\textbf {\bibinfo {volume} {103}},\ \bibinfo {pages} {184202} (\bibinfo {year} {2021})}\BibitemShut {NoStop}%
\bibitem [{\citenamefont {Zerah~Harush}\ and\ \citenamefont {Dubi}(2021)}]{long_range_natural1}%
  \BibitemOpen
  \bibfield  {author} {\bibinfo {author} {\bibfnamefont {E.}~\bibnamefont {Zerah~Harush}}\ and\ \bibinfo {author} {\bibfnamefont {Y.}~\bibnamefont {Dubi}},\ }\href {\doibase https://doi.org/10.1126/sciadv.abc4631} {\bibfield  {journal} {\bibinfo  {journal} {Science advances}\ }\textbf {\bibinfo {volume} {7}},\ \bibinfo {pages} {eabc4631} (\bibinfo {year} {2021})}\BibitemShut {NoStop}%
\bibitem [{\citenamefont {Mattioni}\ \emph {et~al.}(2021)\citenamefont {Mattioni}, \citenamefont {Caycedo-Soler}, \citenamefont {Huelga},\ and\ \citenamefont {Plenio}}]{long_range_artifical1}%
  \BibitemOpen
  \bibfield  {author} {\bibinfo {author} {\bibfnamefont {A.}~\bibnamefont {Mattioni}}, \bibinfo {author} {\bibfnamefont {F.}~\bibnamefont {Caycedo-Soler}}, \bibinfo {author} {\bibfnamefont {S.~F.}\ \bibnamefont {Huelga}}, \ and\ \bibinfo {author} {\bibfnamefont {M.~B.}\ \bibnamefont {Plenio}},\ }\href {\doibase 10.1103/PhysRevX.11.041003} {\bibfield  {journal} {\bibinfo  {journal} {Phys. Rev. X}\ }\textbf {\bibinfo {volume} {11}},\ \bibinfo {pages} {041003} (\bibinfo {year} {2021})}\BibitemShut {NoStop}%
\bibitem [{\citenamefont {Scholes}(2003)}]{long_range_natural2}%
  \BibitemOpen
  \bibfield  {author} {\bibinfo {author} {\bibfnamefont {G.~D.}\ \bibnamefont {Scholes}},\ }\href {\doibase https://doi.org/10.1146/annurev.physchem.54.011002.103746} {\bibfield  {journal} {\bibinfo  {journal} {Annual review of physical chemistry}\ }\textbf {\bibinfo {volume} {54}},\ \bibinfo {pages} {57} (\bibinfo {year} {2003})}\BibitemShut {NoStop}%
\bibitem [{\citenamefont {Browaeys}\ and\ \citenamefont {Lahaye}(2020)}]{browaeys2020many}%
  \BibitemOpen
  \bibfield  {author} {\bibinfo {author} {\bibfnamefont {A.}~\bibnamefont {Browaeys}}\ and\ \bibinfo {author} {\bibfnamefont {T.}~\bibnamefont {Lahaye}},\ }\href {\doibase https://doi.org/10.1038/s41567-019-0733-z} {\bibfield  {journal} {\bibinfo  {journal} {Nature Physics}\ }\textbf {\bibinfo {volume} {16}},\ \bibinfo {pages} {132} (\bibinfo {year} {2020})}\BibitemShut {NoStop}%
\bibitem [{\citenamefont {Joshi}\ \emph {et~al.}(2022)\citenamefont {Joshi}, \citenamefont {Kranzl}, \citenamefont {Schuckert}, \citenamefont {Lovas}, \citenamefont {Maier}, \citenamefont {Blatt}, \citenamefont {Knap},\ and\ \citenamefont {Roos}}]{joshi2022observing}%
  \BibitemOpen
  \bibfield  {author} {\bibinfo {author} {\bibfnamefont {M.~K.}\ \bibnamefont {Joshi}}, \bibinfo {author} {\bibfnamefont {F.}~\bibnamefont {Kranzl}}, \bibinfo {author} {\bibfnamefont {A.}~\bibnamefont {Schuckert}}, \bibinfo {author} {\bibfnamefont {I.}~\bibnamefont {Lovas}}, \bibinfo {author} {\bibfnamefont {C.}~\bibnamefont {Maier}}, \bibinfo {author} {\bibfnamefont {R.}~\bibnamefont {Blatt}}, \bibinfo {author} {\bibfnamefont {M.}~\bibnamefont {Knap}}, \ and\ \bibinfo {author} {\bibfnamefont {C.~F.}\ \bibnamefont {Roos}},\ }\href {\doibase 10.1126/science.abk2400} {\bibfield  {journal} {\bibinfo  {journal} {Science}\ }\textbf {\bibinfo {volume} {376}},\ \bibinfo {pages} {720} (\bibinfo {year} {2022})}\BibitemShut {NoStop}%
\bibitem [{\citenamefont {Porras}\ and\ \citenamefont {Cirac}(2004)}]{porras2004effective}%
  \BibitemOpen
  \bibfield  {author} {\bibinfo {author} {\bibfnamefont {D.}~\bibnamefont {Porras}}\ and\ \bibinfo {author} {\bibfnamefont {J.~I.}\ \bibnamefont {Cirac}},\ }\href {\doibase 10.1103/PhysRevLett.92.207901} {\bibfield  {journal} {\bibinfo  {journal} {Phys. Rev. Lett.}\ }\textbf {\bibinfo {volume} {92}},\ \bibinfo {pages} {207901} (\bibinfo {year} {2004})}\BibitemShut {NoStop}%
\bibitem [{\citenamefont {Defenu}\ \emph {et~al.}(2023)\citenamefont {Defenu}, \citenamefont {Donner}, \citenamefont {Macr\`{\i}}, \citenamefont {Pagano}, \citenamefont {Ruffo},\ and\ \citenamefont {Trombettoni}}]{defenu2023long}%
  \BibitemOpen
  \bibfield  {author} {\bibinfo {author} {\bibfnamefont {N.}~\bibnamefont {Defenu}}, \bibinfo {author} {\bibfnamefont {T.}~\bibnamefont {Donner}}, \bibinfo {author} {\bibfnamefont {T.}~\bibnamefont {Macr\`{\i}}}, \bibinfo {author} {\bibfnamefont {G.}~\bibnamefont {Pagano}}, \bibinfo {author} {\bibfnamefont {S.}~\bibnamefont {Ruffo}}, \ and\ \bibinfo {author} {\bibfnamefont {A.}~\bibnamefont {Trombettoni}},\ }\href {\doibase 10.1103/RevModPhys.95.035002} {\bibfield  {journal} {\bibinfo  {journal} {Rev. Mod. Phys.}\ }\textbf {\bibinfo {volume} {95}},\ \bibinfo {pages} {035002} (\bibinfo {year} {2023})}\BibitemShut {NoStop}%
\bibitem [{\citenamefont {Schuckert}\ \emph {et~al.}(2020{\natexlab{a}})\citenamefont {Schuckert}, \citenamefont {Lovas},\ and\ \citenamefont {Knap}}]{Knap_nonlocal}%
  \BibitemOpen
  \bibfield  {author} {\bibinfo {author} {\bibfnamefont {A.}~\bibnamefont {Schuckert}}, \bibinfo {author} {\bibfnamefont {I.}~\bibnamefont {Lovas}}, \ and\ \bibinfo {author} {\bibfnamefont {M.}~\bibnamefont {Knap}},\ }\href {\doibase 10.1103/PhysRevB.101.020416} {\bibfield  {journal} {\bibinfo  {journal} {Phys. Rev. B}\ }\textbf {\bibinfo {volume} {101}},\ \bibinfo {pages} {020416} (\bibinfo {year} {2020}{\natexlab{a}})}\BibitemShut {NoStop}%
\bibitem [{\citenamefont {Block}\ \emph {et~al.}(2022)\citenamefont {Block}, \citenamefont {Bao}, \citenamefont {Choi}, \citenamefont {Altman},\ and\ \citenamefont {Yao}}]{block_measurement-induced_2022}%
  \BibitemOpen
  \bibfield  {author} {\bibinfo {author} {\bibfnamefont {M.}~\bibnamefont {Block}}, \bibinfo {author} {\bibfnamefont {Y.}~\bibnamefont {Bao}}, \bibinfo {author} {\bibfnamefont {S.}~\bibnamefont {Choi}}, \bibinfo {author} {\bibfnamefont {E.}~\bibnamefont {Altman}}, \ and\ \bibinfo {author} {\bibfnamefont {N.~Y.}\ \bibnamefont {Yao}},\ }\href {\doibase 10.1103/PhysRevLett.128.010604} {\bibfield  {journal} {\bibinfo  {journal} {Phys. Rev. Lett.}\ }\textbf {\bibinfo {volume} {128}},\ \bibinfo {pages} {010604} (\bibinfo {year} {2022})}\BibitemShut {NoStop}%
\bibitem [{\citenamefont {M\"uller}\ \emph {et~al.}(2022)\citenamefont {M\"uller}, \citenamefont {Diehl},\ and\ \citenamefont {Buchhold}}]{buchhold_effective_2021}%
  \BibitemOpen
  \bibfield  {author} {\bibinfo {author} {\bibfnamefont {T.}~\bibnamefont {M\"uller}}, \bibinfo {author} {\bibfnamefont {S.}~\bibnamefont {Diehl}}, \ and\ \bibinfo {author} {\bibfnamefont {M.}~\bibnamefont {Buchhold}},\ }\href {\doibase 10.1103/PhysRevLett.128.010605} {\bibfield  {journal} {\bibinfo  {journal} {Phys. Rev. Lett.}\ }\textbf {\bibinfo {volume} {128}},\ \bibinfo {pages} {010605} (\bibinfo {year} {2022})}\BibitemShut {NoStop}%
\bibitem [{\citenamefont {Minato}\ \emph {et~al.}(2022)\citenamefont {Minato}, \citenamefont {Sugimoto}, \citenamefont {Kuwahara},\ and\ \citenamefont {Saito}}]{Minato_MIPT_long_range}%
  \BibitemOpen
  \bibfield  {author} {\bibinfo {author} {\bibfnamefont {T.}~\bibnamefont {Minato}}, \bibinfo {author} {\bibfnamefont {K.}~\bibnamefont {Sugimoto}}, \bibinfo {author} {\bibfnamefont {T.}~\bibnamefont {Kuwahara}}, \ and\ \bibinfo {author} {\bibfnamefont {K.}~\bibnamefont {Saito}},\ }\href {\doibase 10.1103/PhysRevLett.128.010603} {\bibfield  {journal} {\bibinfo  {journal} {Phys. Rev. Lett.}\ }\textbf {\bibinfo {volume} {128}},\ \bibinfo {pages} {010603} (\bibinfo {year} {2022})}\BibitemShut {NoStop}%
\bibitem [{\citenamefont {Dziarmaga}\ \emph {et~al.}(2012)\citenamefont {Dziarmaga}, \citenamefont {Zurek},\ and\ \citenamefont {Zwolak}}]{dziarmaga_non-local_2012}%
  \BibitemOpen
  \bibfield  {author} {\bibinfo {author} {\bibfnamefont {J.}~\bibnamefont {Dziarmaga}}, \bibinfo {author} {\bibfnamefont {W.~H.}\ \bibnamefont {Zurek}}, \ and\ \bibinfo {author} {\bibfnamefont {M.}~\bibnamefont {Zwolak}},\ }\href {https://www.nature.com/articles/nphys2156} {\bibfield  {journal} {\bibinfo  {journal} {Nat. Phys.}\ }\textbf {\bibinfo {volume} {8}},\ \bibinfo {pages} {49} (\bibinfo {year} {2012})}\BibitemShut {NoStop}%
\bibitem [{\citenamefont {Schuckert}\ \emph {et~al.}(2020{\natexlab{b}})\citenamefont {Schuckert}, \citenamefont {Lovas},\ and\ \citenamefont {Knap}}]{schuckert_nonlocal_2020}%
  \BibitemOpen
  \bibfield  {author} {\bibinfo {author} {\bibfnamefont {A.}~\bibnamefont {Schuckert}}, \bibinfo {author} {\bibfnamefont {I.}~\bibnamefont {Lovas}}, \ and\ \bibinfo {author} {\bibfnamefont {M.}~\bibnamefont {Knap}},\ }\href {https://link.aps.org/doi/10.1103/PhysRevB.101.020416} {\bibfield  {journal} {\bibinfo  {journal} {Phys. Rev. B}\ }\textbf {\bibinfo {volume} {101}},\ \bibinfo {pages} {020416} (\bibinfo {year} {2020}{\natexlab{b}})}\BibitemShut {NoStop}%
\bibitem [{\citenamefont {Dhar}\ \emph {et~al.}(2013)\citenamefont {Dhar}, \citenamefont {Saito},\ and\ \citenamefont {Derrida}}]{dhar_exact_2013}%
  \BibitemOpen
  \bibfield  {author} {\bibinfo {author} {\bibfnamefont {A.}~\bibnamefont {Dhar}}, \bibinfo {author} {\bibfnamefont {K.}~\bibnamefont {Saito}}, \ and\ \bibinfo {author} {\bibfnamefont {B.}~\bibnamefont {Derrida}},\ }\href {https://link.aps.org/doi/10.1103/PhysRevE.87.010103} {\bibfield  {journal} {\bibinfo  {journal} {Phys. Rev. E}\ }\textbf {\bibinfo {volume} {87}},\ \bibinfo {pages} {010103} (\bibinfo {year} {2013})}\BibitemShut {NoStop}%
\bibitem [{\citenamefont {Lorber}\ \emph {et~al.}(2024)\citenamefont {Lorber}, \citenamefont {Zimron}, \citenamefont {Zak}, \citenamefont {Milo},\ and\ \citenamefont {Dubi}}]{lorber_using_2024}%
  \BibitemOpen
  \bibfield  {author} {\bibinfo {author} {\bibfnamefont {S.}~\bibnamefont {Lorber}}, \bibinfo {author} {\bibfnamefont {O.}~\bibnamefont {Zimron}}, \bibinfo {author} {\bibfnamefont {I.~L.}\ \bibnamefont {Zak}}, \bibinfo {author} {\bibfnamefont {A.}~\bibnamefont {Milo}}, \ and\ \bibinfo {author} {\bibfnamefont {Y.}~\bibnamefont {Dubi}},\ }\href {https://link.aps.org/doi/10.1103/PhysRevApplied.22.014041} {\bibfield  {journal} {\bibinfo  {journal} {Physical Review Applied}\ }\textbf {\bibinfo {volume} {22}},\ \bibinfo {pages} {014041} (\bibinfo {year} {2024})}\BibitemShut {NoStop}%
\end{thebibliography}%

\end{document}